\documentclass[12pt]{article}
\usepackage{amsmath,amssymb,epsfig}
\makeatletter \@addtoreset{equation}{section} \makeatother
\renewcommand{\theequation}{\thesection.\arabic{equation}}
\newcommand{\ba}{\begin{array}}
\newcommand{\ea}{\end{array}}
\newcommand{\beq}{\begin{equation}}
\newcommand{\eeq}{\end{equation}}
\newcommand{\bea}{\begin{eqnarray}}
\newcommand{\eea}{\end{eqnarray}}

\def\bce{\begin{center}}
\def\ece{\end{center}}

\def\nonu{\nonumber}

\def\be{\beta}

\newcommand{\tr}{\mbox{Tr}}

\def\eps6{{\displaystyle \mathop{\epsilon}^{6}}{}}

\def\nab6{{\displaystyle \mathop{\nabla}^{6}}{}}

\def\0{{\sst{(0)}}}
\def\1{{\sst{(1)}}}
\def\2{{\sst{(2)}}}
\def\3{{\sst{(3)}}}
\def\4{{\sst{(4)}}}
\def\5{{\sst{(5)}}}
\def\6{{\sst{(6)}}}
\def\7{{\sst{(7)}}}
\def\8{{\sst{(8)}}}

\def\ba{\begin{array}}
\def\ea{\end{array}}
\def\beq{\begin{equation}}
\def\eeq{\end{equation}}
\def\be{\begin{equation}}
\def\ee{\end{equation}}

\def\tr{\mathop{\rm tr}}

\def\eps{\epsilon}

\def\ba{\begin{array}}
\def\ea{\end{array}}
\def\beq{\begin{equation}}
\def\eeq{\end{equation}}
\def\be{\begin{equation}}
\def\ee{\end{equation}}

\def\tr{\mathop{\rm tr}}

\def\eps{\epsilon}

\newcommand{\bean}{\begin{eqnarray*}}
\newcommand{\eean}{\end{eqnarray*}}

\begin{document}
\thispagestyle{empty} \addtocounter{page}{-1}
   \begin{flushright}
KIAS-P08001 \\
\end{flushright}

\vspace*{1.3cm}

\centerline{ \Large \bf  Meta-Stable Brane Configurations }
\vspace{.3cm} 
\centerline{ \Large \bf by Quartic Superpotential for Bifundamentals } 
\vspace*{1.5cm}
\centerline{{\bf Changhyun Ahn} 
} 
\vspace*{1.0cm} 
\centerline{\it 
Department of Physics, Kyungpook National University, Taegu
702-701, Korea} 
\vspace*{0.8cm} 
\centerline{\tt ahn@knu.ac.kr} 
\vskip2cm

\centerline{\bf Abstract}
\vspace*{0.5cm}

The type IIA
nonsupersymmetric meta-stable brane
configuration consisting of three NS5-branes, D4-branes and 
anti-D4-branes where the electric gauge theory superpotential has 
a quartic term for the bifundamentals besides a mass term 
is constructed.
By adding the orientifold 4-plane and 6-plane to this brane
configuration,
we also describe the intersecting brane configurations of type IIA 
string theory corresponding to the meta-stable nonsupersymmetric 
vacua of corresponding gauge theories.  

\baselineskip=18pt
\newpage
\renewcommand{\theequation}
{\arabic{section}\mbox{.}\arabic{equation}}

\section{Introduction}

It has been found that 
the dynamical supersymmetry breaking in meta-stable vacua \cite{ISS,IS} 
can occur
in the ${\cal N}=1$ 
supersymmetric gauge theory with massive fundamental 
flavors.
The extra mass term
for the quarks in the superpotential has led to the fact that
some of the F-term equations cannot  be satisfied and then the
supersymmetry is broken.
The meta-stable brane realizations of type IIA string theory 
have been studied in \cite{OO1,FGU,BGHSS}.  

Recently it has been found in \cite{GK0710-1,GK0710} that other kinds
of 
the type IIA nonsupersymmetric meta-stable 
brane configuration can be constructed by considering  
an additional  quartic
term for the quarks in the superpotential besides the mass term for
the quarks.
Geometrically, this extra deformation in the gauge theory 
corresponds to the rotation of D6-branes 
along the (45)-(89) directions in type IIA string theory realization. 
Classically there exist only
supersymmetric ground states because due to the extra quartic term all
the F-term equations are satisfied. 
By adding the orientifold 6-plane to this brane
configuration \cite{GK0710-1},
the brane configuration 
corresponding to the meta-stable nonsupersymmetric 
vacua of the supersymmetric unitary gauge theory with symmetric flavor
as well as fundamental flavors is found \cite{Ahn07-11}. 
For the antisymmetric flavor case,
the corresponding meta-stable brane configuration is also 
described in \cite{Ahn08-1}.

On the other hand, 
the NS5-brane configuration in type IIA string theory
where 
there exist two types of NS5-branes, i.e., NS5-brane(012345) and 
NS5'-brane(012389), preserves ${\cal N}=2$
supersymmetry in four dimensions \cite{GK98}.
By adding D4-branes(01236) and 
anti-D4-branes($\overline{D4}$-branes) into this system, the
supersymmetry is broken \cite{GK}. 
As the distance between the two NS5'-branes 
becomes zero, this brane configuration with
D4- and $\overline{D4}$-branes can decay and the geometric 
misalignment between 
flavor D4-branes arises.  
Also the meta-stable vacua of \cite{ISS} appear in some
region of parameter space. 

It is natural to ask what happens when some of the NS-branes in the brane
configuration of \cite{GK} are rotated, as suggested in \cite{GK0710-1}? 
Recall that what Giveon and Kutasov did in
 \cite{GK0710-1,GK0710} is to rotate D6-branes with some angles,
compared to the brane configuration of \cite{OO1,FGU,BGHSS}. 
What we do in this paper is to rotate some of the NS-branes with some
angles, compared to  
the brane configuration of \cite{GK}.

One expects, in the gauge theory side, 
that the quartic term for the bifundamentals appears in the superpotential. 
First, we construct the meta-stable brane configuration 
by rotating the NS-brane in the brane configuration of \cite{GK} and 
secondly, we focus on the meta-stable brane configurations 
by adding an orientifold 4-plane and an orientifold 6-plane to 
this brane configuration, along the line of 
\cite{OO1,FGU,BGHSS,Ahn06,Ahn06-1}.
When the former is added, no extra NS5-branes or D-branes are needed.
However, when the latter is added, the extra NS5-branes or D-branes 
into the above
brane configuration are needed in order to have a product gauge group. 
All of these examples have very simple dual magnetic superpotentials 
which make it easier to analyze the meta-stable brane 
configurations.  

In section 2, we review the type IIA brane configuration corresponding
to the electric theory based on the ${\cal N}=1$ $SU(N_c) \times
SU(N_c')$ 
gauge theory 
with the bifundamentals and deform this theory by adding both the mass term
and the quartic term for the bifundamentals.
In the brane configuration, this is equivalent to 
a displacement and a rotation of NS5'-brane. 
Then we construct the dual magnetic theory which is 
${\cal N}=1$ $SU(\widetilde{N}_c) \times SU(N_c')$ gauge 
theory with corresponding dual
matter as well as gauge singlet for the first gauge group factor. 
This corresponds to an interchange of two NS-branes.
We consider the nonsupersymmetric meta-stable
minimum  and present 
the corresponding intersecting brane configurations of type IIA string
theory. Some of the flavor D4-branes are approaching the NS5-brane.

In section 3, we review the type IIA brane configuration corresponding
to the electric theory based on the ${\cal N}=1$ $Sp(N_c) \times
SO(2N_c')$ 
gauge theory 
with a bifundamental and deform this theory by adding the mass term
and the quartic term for the bifundamental. 
Due to the presence of O4-plane, 
a displacement and a rotation of NS5'-brane occur also for the mirror
of NS5'-brane. 
Then we construct the dual magnetic theory which is 
${\cal N}=1$ $Sp(\widetilde{N}_c) \times SO(2N_c')$ gauge 
theory with corresponding dual
matter as well as gauge singlet for the first gauge group factor. 
We consider the nonsupersymmetric meta-stable
minimum  and present 
the corresponding intersecting brane configurations of type IIA string
theory.
Detaching of flavor D4-branes happens also for the mirrors. 
We also discuss  the dual magnetic theory which is 
${\cal N}=1$ $Sp(N_c) \times SO(2\widetilde{N}_c')$ gauge 
theory briefly. Contrary to the unitary case in section 2, 
the rank of gauge group and matter contents are different from
the one in previous case.  

In section 4, we describe the type IIA brane configuration corresponding
to the electric theory based on the ${\cal N}=1$ $SU(N_c) \times
SU(N_c')$ gauge theory 
with  different matters and deform this theory by adding both the mass term
and the quartic term for the bifundamentals. 
Due to the presence of O6-plane, 
a displacement and a rotation of NS5-brane occur also for the mirror
of NS5-brane. 
Then we construct the  dual magnetic theory which is 
${\cal N}=1$ $SU(\widetilde{N}_c) 
\times SU(N_c')$ gauge 
theory with corresponding dual
matters as well as gauge singlet for the first gauge group factor. 
This corresponds to an interchange of two NS5'-branes.
We consider the nonsupersymmetric meta-stable
minimum  and present 
the corresponding intersecting brane configurations of type IIA string
theory. 
Detaching of flavor D4-branes happens for the mirrors but the O6-plane
action behaves differently, compared with the one of O4-plane in section 3. 
We also consider the same gauge theory with other different matters 
and  describe the nonsupersymmetric meta-stable brane configuration
from the  dual magnetic theory which is 
${\cal N}=1$ $SU(\widetilde{N}_c) \times SU(N_c')$ gauge 
theory.

In section 5,  we make some comments for the future directions after
the summary of this paper.  

\section{Meta-stable brane configuration with three NS-branes }

\subsection{Electric theory}

The type IIA brane configuration \cite{BH,AT97} corresponding to 
${\cal N}=1$ supersymmetric gauge theory with
gauge group
\bea
SU(N_c) \times SU(N_c')
\label{gaugeg}
\eea
and 
a bifundamental $X$ in the representation 
$({\bf N_c, \overline{N_c'}})$ and its conjugate field $\widetilde{X}$ 
in the representation $({\bf \overline{N_c}, N_c'})$, 
under the gauge group (\ref{gaugeg}) can be described as follows: 
the middle NS5-brane(012345),
the left $NS5_L'$-brane(012389), the right 
$NS5_R'$-brane(012389), $N_c$-  and $N_c'$-color D4-branes(01236).
We take the arbitrary number of color D4-branes with the constraint 
$N_c' \geq N_c$.
The bifundamentals $X$ and $\widetilde{X}$  correspond to 4-4 
strings connecting 
the $N_c$-color D4-branes with $N_c'$-color D4-branes
\footnote{
See also the relevant works in \cite{ILS,BIWW} for gauge theory
analysis in the context of supersymmetric vacua 
and \cite{Ahn07-3,Ahn07-8,Ahn07-10} for 
nonsupersymmetric meta-stable vacua in the product gauge group 
theory.}.

The middle NS5-brane is located at $x^6=0$ and the $x^6$ 
coordinates for the $NS5_L'$-brane and $NS5_R'$-brane 
are given  by $x^6=-y_2$ and
$x^6=y_1$
respectively, along the line of \cite{GK}. 
The $N_c$ D4-branes 
are suspended between the 
NS5-brane(whose $x^6$ coordinate is given by $x^6=0$) and 
$NS5_R'$-brane(whose $x^6$ coordinate is given by $x^6=y_1$) while 
the $N_c'$ D4-branes 
are suspended between the $NS5_L'$-brane(whose 
$x^6$ coordinate is given by $x^6=-y_2$) and the NS5-brane.

We draw this brane configuration \footnote{This is 
equivalent to the reduced brane
configuration of Figure 1 in \cite{Ahn07-3} if we remove or ignore 
all the D6-branes 
completely.} in Figure 1A for the vanishing mass
for the bifundamentals \cite{GK,Ahn07-3}.
The gauge couplings of $SU(N_c)$ and $ SU(N_c')$
are given by
\bea
g_1^2 = \frac{g_s \ell_s}{y_1}, \qquad 
g_2^2 = \frac{g_s \ell_s}{y_2}.
\label{couplings}
\eea
As $y_2$ goes to the infinity, the $SU(N_c')$ gauge group becomes a
global symmetry and the theory above leads to SQCD with the gauge group
$SU(N_c)$ and $N_c'$ flavors in the fundamental representation.

\begin{figure}[ht]
   \epsfxsize=4.0in 
\centerline{\epsffile{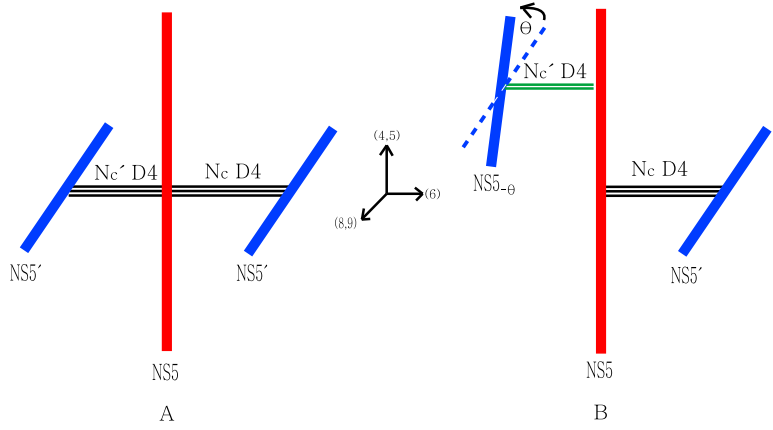}}
   \caption[FIG. \arabic{figure}.]{ 
The  ${\cal N}=1$ supersymmetric 
electric brane configuration for the gauge group $SU(N_c) \times
SU(N_c')$ and the bifundamentals $X$ and $\widetilde{X}$  
with vanishing mass term(1A) and 
nonvanishing mass and quartic terms(1B)
for the bifundamentals. 
In Figure 1B, a 
``rotation'' of $NS5_L'$-brane in $(w,v)$-plane corresponds to 
a quartic term for the bifundamentals while 
a ``displacement'' of $NS5_L'$-brane in $+v$ direction corresponds to a
mass term for the bifundamentals.
The
superpotential for this brane realization of Figure 1B 
is given by (\ref{superelectric}) with the
conditions (\ref{conditions}).}
\end{figure}

According to the result of \cite{Ahn07-3}, there is no electric
superpotential for the Figure 1A. Now let us deform 
this theory. 
Displacing the two NS5'-branes relative each other in the 
\bea
v \equiv x^4 + i x^5
\nonu
\eea
direction \cite{GK98} corresponds to turning on a quadratic
superpotential
for the bifundamentals $X$ and $\widetilde{X}$.
Furthermore, 
rotating the NS5'-branes in the 
$(v,w)$ plane
where we introduce \cite{GK98}
\bea
w \equiv x^8 + i x^9
\nonu
\eea
corresponds to turning on a quartic
superpotential
for the bifundamentals $X$ and $\widetilde{X}$ \cite{BH,BHKL}.
Let us denote them by $NS5_{L,-\theta_1}$-brane and
$NS5_{R,-\theta_2}$-brane
which are at angle $-\theta_1$ and $-\theta_2$ in $(w,v)$-plane
respectively
\footnote{The convention for
$NS5_{-\theta}$-brane here
is different from the one in \cite{Ahn07,Ahn07-1} where the angle between
unrotated NS5'-brane and $NS5_{-\theta}$-brane was not $\theta$ but 
$(\frac{\pi}{2}-\theta)$.}. 
Then the deformed electric superpotential is given by \cite{BH,BHKL}
\bea
W_{elec} = -\frac{\alpha}{2}   \tr (X \widetilde{X})^2 + m \tr X \widetilde{X},
\qquad
\alpha = \frac{1}{\Lambda} \left(\tan \theta_1 + \tan \theta_2 \right), 
\qquad m = \frac{v_{NS5_{-\theta_1}}}{2\pi \ell_s^2}. 
\label{superelectric}
\eea
Here, the $NS5_L'$-brane is moving to the $+v$ direction with $N_c'$
D4-branes 
and 
the $x^5$ coordinate of $NS5_L'$-brane is given by $+ v_{NS5_{-\theta_1}}$.
We focus on the case where 
\bea
\theta_2=0 \qquad \mbox{and} \qquad \theta_1 \equiv \theta.
\label{conditions}
\eea
That is, the $NS5_L'$-brane becomes 
$NS5_{L,-\theta}$-brane and the $NS5_R'$-brane remains  
$NS5_{R,-0}=NS5_R'$-brane under the rotation. 
Giving an expectation value to the meson field $X \widetilde{X}$
corresponds to recombination of $N_c$- and $N_c'$- color 
D4-branes in Figure 1A such that they are suspended between 
the $NS5_L'$-brane and the $NS5_R'$-brane 
and pushing $N_c$ D4-branes into the 
$w$ direction. 

We draw the deformed brane configuration in Figure 1B for nonvanishing mass
for the bifundamentals
by both moving the $NS5_L'$-brane with 
$N_c'$ color D4-branes 
to the $+v$ direction and rotating it 
by an angle $-\theta$ in $(w,v)$-plane as we mentioned.  
Compared with the brane configuration of \cite{GK}, the
difference is coming from the rotation of $NS5_{L}'$-brane. Of course, 
the $\theta=0$  limit for the Figure 1B reduces to the brane
realization 
of \cite{GK}.

The solution for the supersymmetric vacua can be written as 
$X \widetilde{X} =\frac{m}{\alpha}$ through the F-term conditions.
This breaks 
the gauge group $SU(N_c) \times SU(N_c')$ to $SU(N_c-k), SU(N_c'-k)$
and $U(k)$ \cite{BH}.
When the middle NS5-brane moves to $+w$ direction, then the three
NS-branes intersect in three points in $(v,w)$-plane. Then $(N_c-k)$
D4-branes are connecting between the middle NS5-brane and the
$NS5_R'$-brane.
The $(N_c'-k)$
D4-branes are connecting between the $NS5_{-\theta}$-brane and 
the middle NS5-brane. Finally,  $k$
D4-branes are connecting between the $NS5_{-\theta}$-brane and the
$NS5_R'$-brane directly.
The distance from $k$ D4-branes to the middle NS5-brane can be read
off from the trigonometric geometry and it is given by
$w=v_{NS5_{-\theta}} \cot \theta$ \cite{BH}. 
 
\subsection{Magnetic theory}

Starting from the Figure 1B, we apply the Seiberg dual to the 
$SU(N_c)$ factor in (\ref{gaugeg}), and 
the middle NS5-brane and the right
$NS5_R'$-brane are 
interchanged each other. Then the number of color $\widetilde{N}_c$
was given by $\widetilde{N}_c=N_c'-N_c$ connecting the $NS5_R'$-brane
and the NS5-brane from \cite{GK98,Ahn07-3}.
By moving the NS5-brane in Figure 1B to the right all the
way past the $NS5_R'$-brane,
one obtains the Figure 2A.
Before arriving at the Figure 2A, there exists an intermediate 
step where the $N_c'$ D4-branes are connecting between the 
$NS5_{-\theta}$-brane and $NS5_R'$-brane and 
$\widetilde{N}_c$ D4-branes are connecting between $NS5_R'$-brane and   
NS5-brane. 
By introducing $N_c'$ D4-branes and $N_c'$ 
anti-D4-branes  between $NS5_R'$-brane and   
NS5-brane, reconnecting the former with  
the $N_c'$ D4-branes that are connecting between the 
$NS5_{-\theta}$-brane and the $NS5_R'$-brane and moving those 
combined D4-branes
to $+v$-direction, 
one gets the final Figure 2A where we are left with 
$(N_c'-\widetilde{N}_c)$ anti-D4-branes between $NS5_R'$-brane and   
NS5-brane.

\begin{figure}[ht]
   \epsfxsize=4.0in 
\centerline{\epsffile{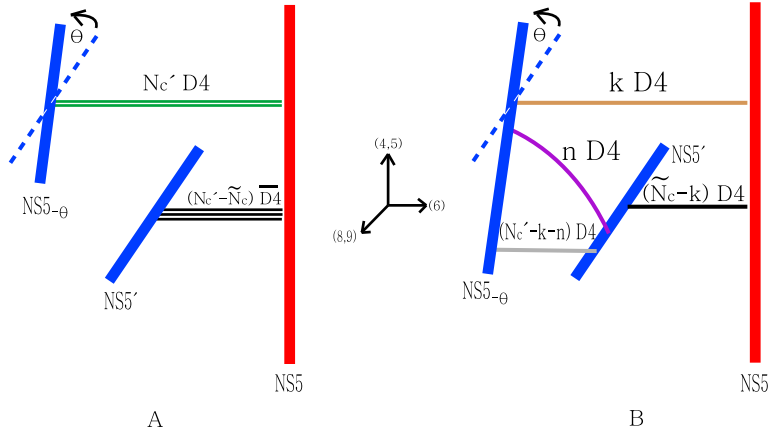}}
   \caption[FIG. \arabic{figure}.]{ 
The  magnetic brane configuration corresponding to Figure 1B with D4-
and $\overline{D4}$-branes(2A) 
when the distance between $NS5_{-\theta}$-brane
and the $NS5_R'$-brane along $v$ direction is large 
and with 
a misalignment between D4-branes(2B) when they are close to
each other. 
At first, the $N_c'$ flavor D4-branes connecting between
$NS5_{-\theta}$-brane and $NS5_R'$-brane are splitting into $(N_c'-k)$ and
$k$ D4-branes. 
The location of intersection between $NS5_{-\theta}$-brane and $(N_c'-k)$
D4-branes 
is given by $(v,w)=(0, +v_{NS5_{-\theta}} 
\cot \theta)$ while the one between  
$NS5_{-\theta}$-brane and $k$
D4-branes 
is given by $(v,w)=(+v_{NS5_{-\theta}}, 0)$. Secondly, 
by moving $n$ flavor D4-branes 
from $(N_c'-k)$ flavor D4-branes,  
the nonzero positive 
$w$ coordinate for $n$ ``curved'' flavor D4-branes is
determined later. }
\end{figure}

The dual gauge group is given by
\bea
SU(\widetilde{N}_c=N_c'-N_c) \times SU(N_c').
\label{dual11}
\eea
The matter contents are the field $Y$ in the bifundamental 
representation $({\bf \widetilde{N}_c, \overline{N_c'}})$,
its complex conjugate field $\widetilde{Y}$ in the bifundamental 
representation  $({\bf \overline{\widetilde{N}_c}, N_c'})$,
under the dual gauge
group (\ref{dual11}) 
and  the gauge singlet $M \equiv X \widetilde{X}$ 
in the representation for 
$({\bf 1,N_c^{'2}-1}) \oplus ({\bf 1, 1})$ under the dual gauge group.
A cubic superpotential arises as an interaction between dual ``quarks''
$Y, \widetilde{Y}$ and a meson $M$. 
The low energy dynamics of the magnetic brane configuration 
can be described by the ${\cal N}=1$ supersymmetric gauge theory
with gauge group (\ref{dual11})
and the gauge couplings for the two gauge group factors are
given by
\bea
g_{1,mag}^2 = \frac{g_s \ell_s}{y_1}, \qquad 
g_{2,mag}^2 = \frac{g_s \ell_s}{y_2-y_1}.
\nonu
\eea

Then the dual magnetic superpotential, by adding the mass term
for the bifundamentals $X, \widetilde{X}$(which can be 
interpreted as a linear term in the meson $M$) and the quartic term 
for the bifundamentals $X, \widetilde{X}$(that is a mass term in the
meson $M$) to the above cubic
superpotential, is given by
\bea
W_{dual} =  \frac{1}{\Lambda} M Y \widetilde{Y}- 
\frac{\alpha}{2} M^2  + m M. 
\label{superpo11}
\eea
The brane configuration for zero mass for the bifundamentals
can be obtained from Figure 2A by  pushing the $NS5_{-\theta}$-brane
together with $N_c'$ D4-branes 
into the origin $v=0$.
Then the number of dual colors for D4-branes 
becomes $N_c'$ between the $NS5_{-\theta}$-brane
and the $NS5_R'$-brane and 
$\widetilde{N}_c$ between the $NS5_R'$-brane and the NS5-brane.
Further zero limit of quartic term for the bifundamentals 
can be achieved by taking $\theta \rightarrow 0$ 
for the $NS5_{-\theta}$-brane 
superpotential (\ref{superpo11}).

The conditions
 $b_{SU(\widetilde{N}_c)}^{mag}=2N_c'-3N_c 
< 0$ and $b_{SU(N_c)}=3N_c-N_c' > 0$ imply
that $N_c' < \frac{3}{2} N_c $. 
Then the range for the $N_c'$ 
can be written as $ N_c < N_c' 
< \frac{3}{2} N_c $. 
Moreover, $b_{SU(N_c')}= 3N_c'-N_c > 0$ and 
$b_{SU(N_c')}^{mag}=N_c+N_c' > 0$.
At the scale $\Lambda_1$, the $SU(N_c)$ theory is strongly
coupled and the Seiberg duality occurs. All the running couplings are
changed by this duality and the coefficients of beta function 
$b_{SU(\widetilde{N}_c)}^{mag}$ becomes negative and 
$b_{SU(N_c')}^{mag}$
becomes positive. 
Then at energy scale lower than $\Lambda_1$, 
the theory is weakly coupled. 
It is not enough 
to choose it lower than Landau pole $\Lambda_1$ simply because one cannot ignore
the contributions from the coupling of $SU(N_c')^{mag}$.
Then under the constraint, $\Lambda_2 <<  \left(
  \frac{\Lambda_1}{\mu}\right)^b
\Lambda_1 << \Lambda_1$
where $b \equiv 
\frac{b_{SU(\widetilde{N}_c)}^{mag}-b_{SU(N_c')}^{mag}}{b_{SU(N_c')}}$, 
one can ignore the contribution from the gauge coupling of 
$SU(N_c')^{mag}$ at the supersymmetry breaking scale and one relies on
the one loop computation. Then one can use the magnetic 
superpotential (\ref{superpo11}) safely.  
See
the appendix B of \cite{AGM07} for the relevant discussions and
 details. 

The brane configuration in Figure 2A is stable as long as the
distance $v_{NS5_{-\theta}}$ between the $NS5_{-\theta}$-brane and 
the $NS5_R'$-brane is large, as
in \cite{GK,GK0710-1}. If they are close to each other, then this brane
configuration is unstable to decay and leads to 
the brane configuration in Figure
2B where some of the flavor D4-branes which are not straight branes 
are approaching to the NS5-brane.
One regards these brane configurations as particular states in the
magnetic gauge theory with the gauge group (\ref{dual11}) and
superpotential (\ref{superpo11}).
When the $NS5_{-\theta}$-brane 
is replaced by $N_c'$ coincident
D6-branes,
the brane configuration of Figure 2B is the same as 
the one studied in \cite{GK0710-1} where the gauge group was
$SU(n_f-n_c)$
and the matter contents were $n_f$ fundamentals and
gauge singlet.
Then the present number $N_c'$ corresponds to $n_f$ while $N_c$
corresponds to $n_c$. This is equivalent to gauge 
the $U(n_f)$ global symmetry of \cite{GK0710-1} in the low energy. 

At first, in order to obtain the supersymmetric vacua, 
one solves the F-term equations for the superpotential (\ref{superpo11}):
\bea
 M Y & = & 0, \qquad \widetilde{Y} M =0, 
\nonu \\
-\frac{1}{\Lambda} Y  \widetilde{Y} & = & m -\alpha M. 
\label{Fterm}
\eea
The last relation, by multiplying $M$ both sides,  
implies the following matrix equation
$
m M = \alpha M^2$.
Since the eigenvalues for the meson field $M$
are either $0$ or $\frac{m}{\alpha}$, one takes
$N_c' \times N_c'$ matrix $M$ with $k$'s eigenvalues $0$ and $(N_c'-k)$'s
eigenvalues $\frac{m}{\alpha}$ as follows:
\bea
M = \left(
\begin{array}{cc}
0 & 0  \\
0 & \frac{m}{\alpha} {\bf 1}_{N_c'-k}
\end{array}
\right)
\label{M0}
\eea
where $k=1, 2, \cdots, N_c'$ and ${\bf 1}_{N_c'-k}$ is the $(N_c'-k) \times
(N_c'-k)$ identity matrix \cite{GK0710}.
In the brane configuration of Figure 2B, the $k$ of the
$N_c'$ flavor D4-branes are connected with $k$ of $\widetilde{N}_c$ color
D4-branes
and the resulting D4-branes stretch from the $NS5_{-\theta}$-brane to
the NS5-brane directly 
and the coordinate of an intersection point between the 
$k$ D4-branes and the NS5-brane is given by $(v, w)=(+v_{NS5_{-\theta}}, 0)$.
This corresponds to  exactly the $k$'s eigenvalues $0$ of 
$M$ above (\ref{M0}).
Now the remaining $(N_c'-k)$ flavor D4-branes between 
the $NS5_{-\theta}$-branes and 
the $NS5_R'$-brane are related to the corresponding remaining eigenvalues 
of $M$ (\ref{M0}), i.e.,   
$\frac{m}{\alpha} {\bf 1}_{N_c'-k}$.
The coordinate of an intersection point between the 
$(N_c'-k)$ D4-branes and the $NS5_R'$-brane is given 
by $(v, w)=(0, +v_{NS5_{-\theta}} \cot \theta)$.

After we substitute (\ref{M0}) into the last equation of (\ref{Fterm})
gives rise to 
\bea
Y  \widetilde{Y} = \left(
\begin{array}{cc}
m \Lambda {\bf 1}_k & 0  \\
0 & 0
\end{array}
\right).
\label{solqq}
\eea
Since the rank of the left hand side of this is at most $\widetilde{N}_c$,
one must have more stringent bound $k \leq \widetilde{N}_c$.
In the $k$-th vacuum the gauge symmetry is broken to $SU(\widetilde{N}_c-k)$
and 
the supersymmetric vacuum drawn in Figure 2B with $k=0$ has 
$Y  =  \widetilde{Y} =0$ and the gauge group 
$SU(\widetilde{N}_c)$ is unbroken. The expectation value of $M$ in this case 
is given by
$M = \frac{m}{\alpha} {\bf 1}_{N_c'}= m \Lambda \cot \theta {\bf 1}_{N_c'}$.


So far, the ground states are supersymmetric. On the other hand,
the theory has many nonsupersymmetric meta-stable ground states.
For the IR free region, $N_c < N_c' < \frac{3}{2} N_c$ \cite{ISS}, 
the magnetic theory is the effective low energy description of the
asymptotically free electric gauge theory.
When we rescale the meson field as
$M = h \Lambda \Phi $,
then the Kahler potential for $\Phi$ is canonical and the magnetic
``quarks'' are canonical near the origin of field space.
Then the magnetic superpotential can be written in terms of $\Phi$
\bea
W_{dual} = 
 h \Phi  Y   \widetilde{Y} 
 +  
\frac{h^2 \mu_{\phi}}{2} \tr \Phi^2- h \mu^2 \tr \Phi. 
\label{Dual}
\eea
From this, one can read off the following quantities
\bea
\mu^2 = -m \Lambda, 
\qquad \mu_{\phi} = -\alpha \Lambda^2, 
\qquad M = h \Lambda \Phi.
\nonu
\eea

The classical supersymmetric vacua given by (\ref{M0}) and
(\ref{solqq})
can be described as 
\bea
 h \Phi
 = \left(
\begin{array}{cc}
0 & 0  \\
0 & \frac{\mu^2}{\mu_{\phi}} {\bf 1}_{N_c'-k}
\end{array}
\right), \qquad
Y  \widetilde{Y} = \left(
\begin{array}{cc}
\mu^2 {\bf 1}_k & 0  \\
0 & 0
\end{array}
\right).
\nonu
\eea
Now one splits, as in \cite{GK0710-1,GK0710}, 
the $(N_c'-k) \times (N_c'-k)$
block  at the lower right corner of $h\Phi$ and $Y
\widetilde{Y}$ 
into blocks of 
size $n$ and $(N_c'-k-n)$ as follows:
\bea
h\Phi = \left(
\begin{array}{ccc}
0 & 0 & 0  \\
0 & h \Phi_n & 0 \\
0 & 0 & \frac{\mu^2}{\mu_{\phi}} {\bf 1}_{N_c'-k-n}
\end{array}
\right), \qquad
Y  \widetilde{Y} = \left(
\begin{array}{ccc}
\mu^2 {\bf 1}_k & 0 & 0  \\
0 & { \varphi}  \widetilde{\varphi}  &  0 \\
0 & 0 & 0
\end{array}
\right).
\label{Eigen}
\eea
Here $\varphi$ and $\widetilde{\varphi}$ are $n \times (\widetilde{N}_c-k)$
dimensional matrices and correspond to $n$ flavors of fundamentals of
the gauge group $SU(\widetilde{N}_c-k)$ which is unbroken by the nonzero
expectation value of $Y$ and $\widetilde{Y}$ (\ref{solqq}).
In the brane configuration from Figure 2B, 
they correspond to 
fundamental strings connecting between the $n$ flavor D4-branes and
$(\widetilde{N}_c-k)$
color D4-branes. Moreover,
the $\Phi_n$ and ${ \varphi} 
\widetilde{\varphi}$
are $n \times n$ matrices.
The supersymmetric ground state corresponds to the vacuum expectation
values by
$h\Phi_n= \frac{\mu^2}{\mu_{\phi}} {\bf 1}_{n}, 
\varphi =0= \widetilde{\varphi}$. 

The full one loop potential for 
$\Phi_n, \varphi, \widetilde{\varphi} $
takes the form in \cite{GK0710-1}
and differentiating this potential with respect to 
$\Phi_n$ and putting $\varphi=0=
\widetilde{\varphi}$, one obtains
\bea
h \Phi_n 
\simeq \frac{ \mu_{\phi}^{\ast}}{\widetilde{N}_c}
{\bf 1}_n \qquad \mbox{or} \qquad
M_n \simeq \frac{\alpha \Lambda^3}{\widetilde{N}_c} {\bf 1}_{n}
\label{vac}
\eea
for real $\mu$ and 
we assume here that 
$\mu_{\phi} << \mu << \Lambda_m$. The vacuum energy $V$ is given by
$V \simeq n |h \mu^2|^2$ and
expanding around this solution, one obtains
the eigenvalues for mass matrix for $\varphi$ and 
$\widetilde{\varphi}$ 
and the vacuum (\ref{vac}) is locally stable.

The $n$ flavor D4-branes of brane configuration
in 
Figure 2B can bend due to the fact that there exists an attractive
gravitational interaction
between those flavor D4-branes and the NS5-brane from the DBI action, by
following the procedure of \cite{GK}. 
The correct choice for the ground state of the system 
depends on the parameters $y_1, y_2$ and $v_{NS5_{-\theta}}$. 

One can move $n$ D4-branes, from $(N_c'-k)$ D4-branes stretched
between the $NS5_R'$-brane and the $NS5_{-\theta}$-brane at $w=+v_{NS5_{-\theta}}
\cot \theta $, to the local minimum of the potential and the end
points of these $n$ D4-branes are at a nonzero $w$ 
as in Figure 2B \cite{GK0710-1}.
The remaining $(N_c'-k-n)$ flavor D4-branes between 
the $NS5_{-\theta}$-brane and 
the $NS5_R'$-brane are related to the corresponding eigenvalues 
of $h\Phi$ (\ref{Eigen}), i.e.,  
$\frac{\mu^2}{\mu_{\phi}} {\bf 1}_{N_c'-k-n}$.
The coordinate of an intersection point between the 
$(N_c'-k-n)$ D4-branes and the $NS5_R'$-brane is given 
by $(v, w)=(0, +v_{NS5_{-\theta}} \cot \theta)$.
As we mentioned,
the $k$ D4-branes stretching from the $NS5_{-\theta}$-brane to
the NS5-brane 
correspond to  exactly the $k$'s eigenvalues $0$ of $h\Phi$ (\ref{Eigen}).
Finally, 
the remnant $n$ ``curved'' flavor D4-branes between 
the $NS5_{-\theta}$-branes and 
the $NS5_R'$-brane are related to the corresponding eigenvalues 
(\ref{vac}) 
of $h\Phi_n$. 
Since the eigenvalues of (\ref{vac}) are much smaller than 
$\frac{\mu^2}{\mu_{\phi}} {\bf 1}_{N_c'-k-n}$ of 
(\ref{Eigen}), in Figure 2B, the $n$ curved flavor D4-branes, instead
of $(N_c'-k-n)$ flavor D4-branes,
are nearer to the NS5-brane.   
By explicit computation as in \cite{GK0710-1}, it can be
shown that 
the local minimum occurs at $w \simeq \tan \theta \frac{y^4}{\ell_s^2 
v_{NS5_{-\theta}}}$ with $x^6 \equiv y$.

Therefore, the classical brane construction can generalize 
the gauge theory discussion to the regime where the angle $\theta$ is
of order one and different length parameters are of order $\ell_s$ or
larger. Note that the gauge theory analysis is valid only when 
$\theta$ and $\frac{v_{NS5_{-\theta}}}{\ell_s}$ are much smaller than 
$\ell_s$ \cite{GK0710-1}.

\section{Meta-stable brane configuration with three NS-branes plus O4-plane}

\subsection{Electric theory}

The type IIA brane configuration \cite{Ahn07-5} corresponding to 
${\cal N}=1$ supersymmetric gauge theory with
gauge group
\bea
Sp(N_c) \times SO(2N_c')
\label{gaugegroup}
\eea
and a bifundamental $X$ that is in the representation 
$({\bf 2N_c, 2N_c'})$ under the gauge group (\ref{gaugegroup}) 
can be described by 
a middle NS5-brane(012345),
the left $NS5_L'$-brane(012389) and the right 
$NS5_R'$-brane(012389), $2N_c$-  and $2N_c'$-color D4-branes(01236) as
well as an 
O4-plane(01236). 
We take the arbitrary number of color D4-branes with the constraint 
$N_c' \geq N_c+2$.
The O4-plane acts as $(x^4,x^5,x^7,x^8,x^9) \rightarrow
(-x^4,-x^5,-x^7,
-x^8,-x^9)$ as usual.
The bifundamental $X$  corresponds to 4-4 strings connecting 
the $2N_c$-color D4-branes with $2N_c'$-color D4-branes.

The middle NS5-brane is located at $x^6=0$ and the $x^6$ 
coordinates for the $NS5_L'$-brane and $NS5_R'$-brane 
by $x^6=-y_2$ and
$x^6=y_1$
respectively. 
The $2N_c$ D4-branes and $O4^{+}$-plane 
are suspended between the middle
NS5-brane and $NS5_R'$-brane while 
the $2N_c'$ D4-branes and $O4^{-}$-plane 
are suspended between the $NS5_L'$-brane and the middle NS5-brane.
We draw this brane configuration in Figure 3A \cite{Ahn07-5} 
for the vanishing mass
for the bifundamental $X$
\footnote{This is equivalent to the reduced brane
realization of Figure 1 in \cite{Ahn07-2} if we remove D6-branes 
completely. See also 
the relevant works appeared in 
\cite{Tatar,Ahn97,Hashiba} for supersymmetric vacua and 
\cite{Ahn07-2,Ahn07-8,Ahn07-10} for 
nonsupersymmetric vacua in the product gauge group between symplectic
and orthogonal gauge groups. }. 

\begin{figure}[ht]
   \epsfxsize=4.0in 
\centerline{\epsffile{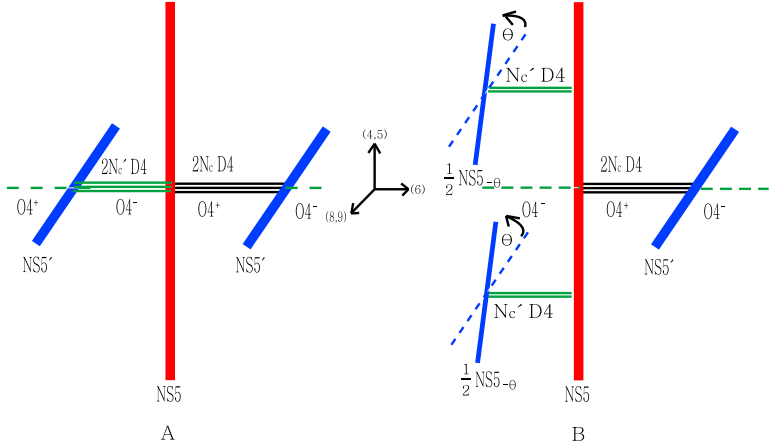}}
   \caption[FIG. \arabic{figure}.]{ 
The  ${\cal N}=1$ supersymmetric 
electric brane configuration for the gauge group $Sp(N_c) \times
SO(2N_c')$ and a bifundamental $X$  with vanishing mass term(3A) and 
nonvanishing mass and quartic terms(3B) 
for the bifundamental. 
In Figure 3B, there are two deformations by  
rotation  and displacement of $NS5_L'$-brane and
the
superpotential is given by (\ref{superelectric1}). Note that the
mirrors for upper half $NS5_{-\theta}$-brane and upper $N_c'$ D4-branes 
in Figure 3B are preserved under the O4-plane
action.}
\end{figure}

The gauge couplings of $Sp(N_c)$ and $ SO(2N_c')$
are given by 
\bea
g_{Sp}^2 = \frac{g_s \ell_s}{y_1}, \qquad 
g_{SO}^2 = \frac{g_s \ell_s}{y_2}
\nonu
\eea
respectively.
As $y_2$ goes to the infinity, the $SO(2N_c')$ gauge group becomes a
global symmetry and the theory leads to SQCD with the gauge group
$Sp(N_c)$ and $N_c'$ flavors(or $2N_c'$ fields) 
in the vector representation.
The opposite limit $y_1 \rightarrow \infty$ leads to SQCD with the
gauge group $SO(2N_c')$ with $2N_c$ fields in the fundamental
representation.

There is no superpotential in Figure 3A.
Let us deform this gauge theory. 
Displacing the two NS5'-branes relative each other in the $v$ 
direction corresponds to turning on a quadratic
mass-deformed superpotential
for the bifundamental $X$  
while rotating them in the $w$ direction corresponds to 
turning on a quartic term for the bifundamental $X$.
The deformed electric superpotential is 
as follows:
\bea
W_{elec} = -\frac{\alpha}{2}   \tr (X X)^2 + m \tr X X,
\qquad
\alpha = \frac{\tan \theta}{\Lambda}, 
\qquad m = \frac{v_{NS5_{-\theta}}}{2\pi \ell_s^2} 
\label{superelectric1}
\eea
where a symplectic metric(that has antisymmetric color 
indices) \cite{Ahn07-2} 
is assumed in the $Sp(N_c)$ gauge group 
indices for a meson field $X X$.
Half of $NS5_{-\theta}$-brane with $N_c'$ color D4-branes 
is moving to the $+v$ direction while the other half of 
$NS5_{-\theta}$-brane  with other $N_c'$ color D4-branes
is moving to $-v$ direction due to the O4-plane for
fixed $NS5_R'$-brane. 
Then the $x^5$ coordinate 
of $NS5_R'$-brane is zero
and the $x^5$ coordinates of each half $NS5_{-\theta}$-brane 
are given by 
$\pm v_{NS5_{-\theta}}$ respectively.

Giving an expectation value to the meson field $X X$
corresponds to recombination of $2N_c$- and $2N_c'$- color 
D4-branes, which becomes $2N_c$-color D4-branes,
in Figure 3A such that they are suspended between 
the $NS5_{-\theta}$-brane and the $NS5_R'$-brane 
and pushing them into the $w$ direction. 

Now 
we draw this brane configuration in Figure 3B for nonvanishing mass
for the bifundamental $X$ by moving half of $NS5_{-\theta}$-brane with 
$N_c'$ color D4-branes to the $+v$ direction and rotating it by an
angle $-\theta$ in $(w,v)$-plane(and their mirrors). 
One can easily understand this brane configuration by adding 
O4-plane into the Figure 1B with appropriate number of color D4-branes.
Compared with the brane configuration of \cite{Ahn07-2,Ahn07-5}, the
difference is the fact that there exists an extra rotation of
$NS5_{-\theta}$-brane. 
Of course, the $\theta=0$ limit reduces to the one of \cite{Ahn07-2,Ahn07-5}.

The solution for the supersymmetric vacua can be obtained by 
$X X =\frac{m}{\alpha}$ through the F-term condition for the 
superpotential (\ref{superelectric1}).
This breaks 
the gauge group $Sp(N_c) \times SO(2N_c')$ to $Sp(N_c-k), SO(2N_c'-2k)$
and $U(2k)$.
When the middle NS5-brane moves to $\pm w$ direction(half of them to
$+w$ direction and half of them to $-w$ direction), then the three
NS-branes($NS5_{-\theta}$-brane, $NS5_R'$-brane and NS5-brane) 
intersect in three points in $(v,w)$-plane. 
In other words, the coordinates of $(v,w)$ for those points are
$(+v_{NS5_{-\theta}}, +v_{NS5_{-\theta}} \cot \theta)$, $(0, 
+v_{NS5_{-\theta}} \cot \theta)$ and $(0, 
+2v_{NS5_{-\theta}} \cot \theta)$.
It is easy to see that the other intersection points 
are given by $(\pm v_{NS5_{-\theta}}, -v_{NS5_{-\theta}} \cot \theta)$, $(0, 
 -v_{NS5_{-\theta}} \cot \theta)$, $(0, 0)$, $(-v_{NS5_{-\theta}}, 
+v_{NS5_{-\theta}} \cot \theta)$ and $(0, 
-2v_{NS5_{-\theta}} \cot \theta)$.
Then $2(N_c-k)$
D4-branes are connecting between the middle NS5-brane and the
$NS5_R'$-brane.
The $2(N_c'-k)$
D4-branes are connecting between the $NS5_{-\theta}$-brane and 
the middle NS5-brane. Finally,  $2k$
D4-branes are connecting between the $NS5_{-\theta}$-brane and the
$NS5_R'$-brane directly.
The distance from $2k$ D4-branes to the middle NS5-brane can be read
off from the trigonometric geometry and the $w$ coordinate is given by
$w=\pm v_{NS5_{-\theta}} \cot \theta$. 

\subsection{Magnetic theory}

By applying the Seiberg dual to the $Sp(N_c)$ factor in 
(\ref{gaugegroup}), starting from Figure 3B and 
moving the NS5-brane to the right all the
way past the $NS5_R'$-brane,
one obtains the Figure 4A.
Before arriving at the Figure 4A, there exists an intermediate 
step where the $N_c'$ D4-branes are connecting between half 
$NS5_{-\theta}$-brane and $NS5_R'$-brane(and their mirrors) and 
$2\widetilde{N}_c$ D4-branes connecting between $NS5_R'$-brane and   
NS5-brane. By introducing $2N_c'$ D4-branes and $2N_c'$ 
anti-D4-branes  between $NS5_R'$-brane and   
NS5-brane, recombining half of the former with  
the $N_c'$ D4-branes that are connecting between half 
$NS5_{-\theta}$-brane and $NS5_R'$-brane and moving those combined D4-branes
to $+v$-direction(and their mirrors), 
one gets the final Figure 4A where we are left with 
$2(N_c'-\widetilde{N}_c)$ anti-D4-branes between $NS5_R'$-brane and   
NS5-brane.

\begin{figure}[ht]
   \epsfxsize=4.0in 
\centerline{\epsffile{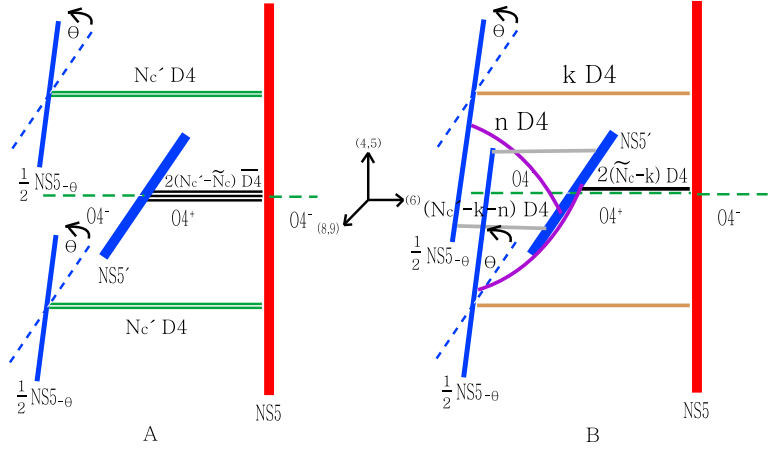}}
   \caption[FIG. \arabic{figure}.]{ 
The  magnetic brane configuration corresponding to Figure 3B with D4-
and $\overline{D4}$-branes(4A) when
the distance between $NS5_{-\theta}$-brane
and the $NS5_R'$-brane along $v$ direction is large 
and with 
a misalignment between D4-branes(4B) when they are close to
each other. 
The upper $N_c'$ flavor D4-branes connecting between
the upper half 
$NS5_{-\theta}$-brane and $NS5_R'$-brane are splitting into $(N_c'-k)$ and
$k$ D4-branes. 
The location of intersection between the upper half $NS5_{-\theta}$-brane
and the upper $(N_c'-k)$
D4-branes 
is given by $(v,w)=(0, +v_{NS5_{-\theta}} 
\cot \theta)$ while the one between  
the upper half $NS5_{-\theta}$-brane and the upper $k$
D4-branes 
is given by $(v,w)=(+v_{NS5_{-\theta}}, 0)$. By moving $n$ flavor D4-branes 
from the upper $(N_c'-k)$ flavor D4-branes,  
the nonzero positive $w$ coordinate for $n$  ``curved'' flavor D4-branes is
determined later.
Similarly, the location of 
intersection between the lower half $NS5_{-\theta}$-brane
and the lower $(N_c'-k)$
D4-branes 
is given by $(v,w)=(0, -v_{NS5_{-\theta}} 
\cot \theta)$ while the one between  
the lower half $NS5_{-\theta}$-brane and the lower $k$
D4-branes 
is given by $(v,w)=(-v_{NS5_{-\theta}}, 0)$. }
\end{figure}

Then the gauge group is given by
\bea
Sp(\widetilde{N}_c=N_c'-N_c-2) \times SO(2N_c')
\label{dualgauge}
\eea
where the number of dual color was obtained from the linking number 
counting, as done in \cite{Ahn07-2,Ahn07-5}.
The matter contents are the field $Y$ in the bifundamental 
representation $({\bf 2\widetilde{N}_c, 2N_c'})$ under the dual gauge
group (\ref{dualgauge}) 
and  the gauge-singlet $M(\equiv X X)$ is
in the adjoint representation for 
the second dual gauge group
$({\bf 1, N_c'(2N_c'-1)})$ under 
the dual gauge group (\ref{dualgauge}).
The quantum corrections can be understood for small $v_{NS5_{-\theta}}$ by 
using the low energy field theory on the branes.
The low energy dynamics of the magnetic brane configuration 
can be described by the ${\cal N}=1$ supersymmetric gauge theory
with gauge group (\ref{dualgauge})
and the gauge couplings for the two gauge group factors are
given by
\bea
g_{Sp,mag}^2 = \frac{g_s \ell_s}{y_1}, \qquad 
g_{SO,mag}^2 = \frac{g_s \ell_s}{(y_2-y_1)}.
\nonu
\eea

Then the dual magnetic superpotential, by adding the mass term
(\ref{superelectric1}) for the bifundamental $X$, which can be 
interpreted as a linear term in the meson $M$, and quartic term 
to the cubic
superpotential, is given by
\bea
W_{dual} =  \frac{1}{\Lambda} M Y Y- \frac{\alpha}{2} M^2  + m M. 
\label{superpo}
\eea
Of course, the brane configuration for zero mass for the bifundamental
can be obtained from Figure 4A by recombination between half
$NS5_{-\theta}$-branes together with color D4-branes via pushing 
them into the origin $v=0$.
Then the number of dual colors for D4-branes 
becomes $2N_c'$ between the $NS5_{-\theta}$-brane and the  
$NS5_R'$-brane 
and $2\widetilde{N}_c$ between $NS5_R'$-brane and NS5-brane.
Moreover, the zero limit of quartic term for the bifundamental
can be done by taking $\theta \rightarrow 0$ for the
$NS5_{-\theta}$-branes.

The conditions
 $b_{Sp(\widetilde{N}_c)}^{mag}=4N_c'-6N_c-6 
< 0$ and $b_{Sp(N_c)}=3(2N_c+2)-2N_c' > 0$ lead to
$N_c' < \frac{3}{2} N_c +\frac{3}{2}$. 
The range for the $N_c'$ 
can be written as $ N_c +2 < N_c' 
< \frac{3}{2} N_c +\frac{3}{2} $. 
Moreover, $b_{SO(2N_c')}= 3(2N_c'-2)-2N_c > 0$ and 
$b_{SO(2N_c')}^{mag}=2(N_c+N_c') > 0$.
At the scale $\Lambda_1$, the $Sp(N_c)$ theory is strongly
coupled and the Seiberg duality occurs. 
Then at energy scale lower than $\Lambda_1$, 
the theory is weakly coupled. 
One cannot ignore
the contributions from the coupling of $SO(2N_c')^{mag}$.
Then under the constraint, $\Lambda_2 <<  \left(
  \frac{\Lambda_1}{\mu}\right)^b
\Lambda_1 << \Lambda_1$
where $b \equiv 
\frac{b_{Sp(\widetilde{N}_c)}^{mag}-b_{SO(2N_c')}^{mag}}{b_{SO(2N_c')}}$, 
one can ignore the contribution from the gauge coupling of 
$SO(2N_c')^{mag}$ at the supersymmetry breaking scale.

The brane configuration in Figure 4A is stable as long as the
distance $v_{NS5_{-\theta}}$ between the upper half $NS5_{-\theta}$-brane and
$NS5_R'$-brane is large, as
in \cite{GK,Ahn07-5}. If they are close to each other, then this brane
configuration is unstable to decay to the brane configuration in Figure
4B with bending effect of tilted 
D4-branes connecting half $NS5_{-\theta}$-brane and $NS5_R'$-brane.
Of course, this brane realization can be obtained from the Figure 
2B by adding O4-plane with appropriate mirrors.
One can regard these brane configurations as particular states in the
magnetic gauge theory with the gauge group (\ref{dualgauge}) and
superpotential (\ref{superpo}). 
When the two half $NS5_{-\theta}$-branes are replaced by the coincident
$2N_c'$ D6-branes,
the brane configuration of Figure 4B is the same as the one studied in 
\cite{GK0710-1} together with an addition of appropriate O4-plane.

In order to obtain the supersymmetric vacua, 
one solves the F-term equations for the superpotential (\ref{superpo}):
\bea
M Y  =  0, \qquad
-\frac{1}{\Lambda} Y  Y  =  m -\alpha M. 
\label{Fterm1}
\eea
The matrix equation
$
m M = \alpha M^2$ implies
that the eigenvalues for the meson field $M$ 
are either $0$ or $\frac{m}{\alpha}$, one takes
$2N_c' \times 2N_c'$ matrix with $2k$'s eigenvalues $0$ and $2(N_c'-k)$'s
eigenvalues $\frac{m}{\alpha}$:
\bea
M = \left(
\begin{array}{cc}
0 & 0  \\
0 & \frac{m}{\alpha} {\bf 1}_{N_c'-k} \otimes i \sigma_2
\end{array}
\right)
\label{M01}
\eea
where $k=1, 2, \cdots, 2N_c'$ and ${\bf 1}_{N_c'-k}$ is the $(N_c'-k) \times
(N_c'-k)$ identity matrix \cite{GK0710} \footnote{The mass matrix $m$
is antisymmetric in the indices and is given by $m =
\mbox{diag}(i\sigma_2 m_1, i\sigma_2 m_2, \cdots, i\sigma_2
m_{N_c'})$ due to the antisymmetric matrix $M$. In the matrix equation
$m M$, we assumed this property of mass matrix. In (\ref{M01}), we
used the equal mass as $m \equiv m_1 = m_2= \cdots = m_{N_c'}$ unfortunately.}. 
Therefore, in the brane configuration of Figure 4B, the $k$ of the
upper $N_c'$ flavor D4-branes are connected with $k$ of $\widetilde{N}_c$ color
D4-branes
and the resulting D4-branes stretch from the upper $NS5_{-\theta}$-brane to
the NS5-brane directly and the coordinate of an intersection point between the 
$k$ upper D4-branes and the NS5-brane is given by $(v,
w)=(+v_{NS5_{-\theta}}, 0)$.
Similarly the mirrors are located at  $(v,
w)=(-v_{NS5_{-\theta}}, 0)$.
This corresponds to  exactly the $k$'s eigenvalues 
$0$ of $M$ above (\ref{M01}).
Now the remaining $(N_c'-k)$ upper flavor D4-branes between 
the $NS5_{-\theta}$-branes and 
the $NS5_R'$-brane are related to the corresponding half eigenvalues 
of $M$ which is equal to  
$\frac{m}{\alpha} {\bf 1}_{N_c'-k} \otimes i \sigma_2$.
The coordinate of an intersection point between the 
$(N_c'-k)$ upper D4-branes and the $NS5_R'$-brane is given 
by $(v, w)=(0, +v_{NS5_{-\theta}} \cot \theta)$ corresponding to
positive eigenvalues of $M$.
The mirrors are located at  
$(v, w)=(0, -v_{NS5_{-\theta}} \cot \theta)$ corresponding to negative
eigenvalues of $M$.

After we substitute (\ref{M01}) into the second equation of (\ref{Fterm1})
gives rise to 
\bea
Y  Y = \left(
\begin{array}{cc}
m \Lambda {\bf 1}_{2k} & 0  \\
0 & 0
\end{array}
\right).
\label{solqq1}
\eea
Since the rank of the left hand side of this is at most $2\widetilde{N}_c$,
one must have more stringent bound $k \leq 2\widetilde{N}_c$.
In the $k$-th vacuum the gauge symmetry is broken to $Sp(\widetilde{N}_c-k)$
and 
the supersymmetric vacuum drawn in Figure 4B with $k=0$ has 
$Y=0$ and the gauge group 
$Sp(\widetilde{N}_c)$ is unbroken. The expectation value of $M$
(\ref{M01}) 
in this case 
is given by
$M = \frac{m}{\alpha} {\bf 1}_{N_c'} \otimes i \sigma_2
= m \Lambda \cot \theta {\bf 1}_{N_c'} \otimes i \sigma_2$.


The theory has many nonsupersymmetric meta-stable ground states.
For the IR free region, $N_c+2 < N_c' < \frac{3}{2} 
(N_c +1)$ \cite{ISS,Ahn06-1}, 
the magnetic theory is the effective low energy description of the
asymptotically free electric gauge theory.
When we rescale the meson field as
$M = h \Lambda \Phi $,
then the Kahler potential for $\Phi$ is canonical and the magnetic
``quarks'' are canonical near the origin of field space.
Then the magnetic superpotential can be written in terms of $\Phi$
\bea
W_{dual} = 
 h \Phi  Y   Y 
 +  
\frac{h^2 \mu_{\phi}}{2} \tr \Phi^2- h \mu^2 \tr \Phi. 
\nonu
\eea

The classical supersymmetric vacua given by (\ref{M01}) and
(\ref{solqq1})
can be described as 
\bea
 h \Phi
 = \left(
\begin{array}{cc}
0 & 0  \\
0 & \frac{\mu^2}{\mu_{\phi}} {\bf 1}_{N_c'-k} \otimes i \sigma_2
\end{array}
\right), \qquad
Y  Y = \left(
\begin{array}{cc}
\mu^2 {\bf 1}_{2k} & 0  \\
0 & 0
\end{array}
\right).
\nonu
\eea
Now one splits, as in \cite{GK0710-1,GK0710}, 
the $2(N_c'-k) \times 2(N_c'-k)$
block  at the lower right corner of $h\Phi$ and $Y Y$ 
into blocks of 
size $2n$ and $2(N_c'-k-n)$ as follows:
\bea
h\Phi = \left(
\begin{array}{ccc}
0 & 0 & 0  \\
0 & h \Phi_{2n}  & 0 \\
0 & 0 & \frac{\mu^2}{\mu_{\phi}} {\bf 1}_{N_c'-k-n} \otimes i \sigma_2
\end{array}
\right), \qquad
Y  Y = \left(
\begin{array}{ccc}
\mu^2 {\bf 1}_{2k} & 0 & 0  \\
0 &  \varphi  \varphi  &  0 \\
0 & 0 & 0
\end{array}
\right).
\label{eigenvalue}
\eea
Here $\varphi$  is $2n \times 2(\widetilde{N}_c-k)$
dimensional matrix and corresponds to $2n$ flavors of fundamentals of
the gauge group $Sp(\widetilde{N}_c-k)$ which is unbroken by the nonzero
expectation value of $Y$.
In the brane configuration in Figure 4B, 
they correspond to 
fundamental strings connecting the $n$ upper flavor D4-branes and
$(\widetilde{N}_c-k)$
color D4-branes(and their mirrors).
The $\Phi_{2n}$ and $ \varphi \varphi$
are $2n \times 2n$ matrices.
The supersymmetric ground state corresponds to
the vacuum expectation values by
$h\Phi_{2n}= \frac{\mu^2}{\mu_{\phi}} {\bf 1}_{n} \otimes i \sigma_2, 
\varphi =0$. 

The full one loop potential for 
$\Phi_{2n}, \varphi$
takes the similar form in \cite{GK0710-1}
and differentiating this potential with respect to 
$\Phi_{2n}$ and putting $\varphi=0$, one obtains
\bea
h \Phi_{2n} 
\simeq \frac{ \mu_{\phi}^{\ast}}{\widetilde{N}_c}
{\bf 1}_n  \otimes i \sigma_2 \qquad \mbox{or} \qquad
M_{2n} \simeq \frac{\alpha \Lambda^3}{\widetilde{N}_c} 
{\bf 1}_{n}  \otimes i \sigma_2
\label{vac1}
\eea
for real $\mu$ and 
we assume here that 
$\mu_{\phi} << \mu << \Lambda_m$. The vacuum energy $V$ is given by
$V \simeq n |h \mu^2|^2$ and
expanding around this solution, one obtains
the eigenvalues for mass matrix for $\varphi$  
and the vacuum (\ref{vac1}) is locally stable.

The $n$ flavor D4-branes of 
straight brane configuration
of
Figure 4B can bend due to the fact that there exists an attractive
gravitational interaction
between those flavor D4-branes and the NS5-brane from the DBI action, by
following the procedure of \cite{GK,Ahn07-5}. 
The correct choice for the ground state of the system 
depends on the parameters $y_1, y_2$ and $v_{NS5_{-\theta}}$. 

One can move $n$ upper D4-branes, from upper $(N_c'-k)$ D4-branes stretched
between the $NS5_R'$-brane and the upper $NS5_{-\theta}$-brane 
at $w=+v_{NS5_{-\theta}}
\cot \theta $, to the local minimum of the potential and the end
points of these $n$ D4-branes are at a nonzero $w$ as in Figure 4B.
The remaining upper $(N_c'-k-n)$ flavor D4-branes between 
the upper $NS5_{-\theta}$-brane and 
the $NS5_R'$-brane are related to the corresponding ``positive'' eigenvalues 
of $h\Phi$ (\ref{eigenvalue}) which is equal to 
$\frac{\mu^2}{\mu_{\phi}} {\bf 1}_{(N_c'-k-n)} 
\otimes i \sigma_2$.
The coordinate of an intersection point between the 
upper $(N_c'-k-n)$ D4-branes and the $NS5_R'$-brane is given 
by $(v, w)=(0, +v_{NS5_{-\theta}} \cot \theta)$.
The remnant $n$ upper ``curved'' flavor D4-branes between 
the $NS5_{-\theta}$-branes and 
the $NS5_R'$-brane are related to the corresponding ``positive'' eigenvalues 
(\ref{vac1}) of $h\Phi_{2n}$.
As we mentioned,
the $k$ D4-branes stretching from the $NS5_{-\theta}$-brane to
the NS5-brane 
correspond to  exactly the $k$'s eigenvalues $0$ of $h\Phi$ (\ref{eigenvalue}).
By explicit computation it can be
shown that 
the local minimum occurs at $w \simeq \tan \theta \frac{y^4}{\ell_s^2 
v_{NS5_{-\theta}}}$ with $x^6 \equiv y$.

Note that the intersection point between the lower $(N_c'-k)$
D4-branes 
and the lower  half $NS5_{-\theta}$-brane 
is located at $w=- v_{NS5_{-\theta}}
\cot \theta$.
The lower $(N_c'-k-n)$ flavor D4-branes 
are related to the corresponding ``negative'' eigenvalues 
of $h\Phi$ (\ref{eigenvalue}).

\subsection{Other magnetic theory }

By applying the Seiberg dual to the $SO(2N_c')$ factor in 
(\ref{gaugegroup}), 
starting from modified Figure 3B, where 
the $x^5$ coordinate 
of $NS5_L'$-brane is equal to
zero
and the $x^5$ coordinates of half $NS5_{-\theta}$-branes which were $NS5_R'$-brane
are $\pm v_{NS5_{-\theta}}$,
and moving the NS5-brane to the left all the
way past the $NS5_L'$-brane,
one obtains the magnetic brane configuration similar to Figure 4A.
The gauge group is given by
\bea
Sp(N_c) 
\times SO(2\widetilde{N}_c'=2N_c-2N_c'+4).
\label{gaugedual}
\eea
The matter contents are the field $Y$ in the bifundamental 
representation $({\bf 2N_c, 2\widetilde{N}_c'})$ under the dual gauge
group (\ref{gaugedual}) 
and  the gauge-singlet $M$ is
in the adjoint representation for 
the first dual gauge group, i.e., a symmetric matrix,
$({\bf N_c(2N_c+1), 1})$ under the dual gauge group.
The superpotential is the same as the one in (\ref{superpo}) and the
corresponding modified 
Figure 4B, which is exactly a reflection of Figure 4B with
respect to the NS5-brane, i.e., all the D4-branes,
$NS5_{-\theta}$-brane 
and $NS5_R'$-brane are
located at the right hand side of NS5-brane, 
can be constructed similarly.
The discussion for the supersymmetric vacua in previous subsection 
can be applied here also  \footnote{In this case, the mass matrix $m$
is symmetric in the indices and is given by $m =
\mbox{diag}(\sigma_3 m_1, \sigma_3 m_2, \cdots, \sigma_3
m_{N_c})$ due to the symmetric matrix $M$. In the matrix equation
$m M$, we assumed this property of mass matrix.}.

The theory has many nonsupersymmetric meta-stable ground states.
For the IR free region, $2N_c'-4 < 2N_c < 
\frac{3}{2} (2N_c' -2)$ \cite{ISS,Ahn06-1}, 
the magnetic theory is the effective low energy description of the
asymptotically free electric gauge theory.
When we rescale the meson field as
$M = h \Lambda \Phi $,
then the Kahler potential for $\Phi$ is canonical and the magnetic
``quarks'' are canonical near the origin of field space.
Then the magnetic superpotential 
can be written in terms of $\Phi$
\bea
W_{dual} = 
 h \Phi  Y   Y 
 +  
\frac{h^2 \mu_{\phi}}{2} \tr \Phi^2- h \mu^2 \tr \Phi. 
\nonu
\eea

The classical supersymmetric vacua 
can be described as 
\bea
 h \Phi
 = \left(
\begin{array}{cc}
0 & 0  \\
0 & \frac{\mu^2}{\mu_{\phi}} {\bf 1}_{N_c-k}  \otimes  \sigma_3
\end{array}
\right), \qquad
Y  Y = \left(
\begin{array}{cc}
\mu^2 {\bf 1}_{2k} & 0  \\
0 & 0
\end{array}
\right).
\nonu
\eea
Now one splits, as in \cite{GK0710-1,GK0710}, 
the $2(N_c-k) \times 2(N_c-k)$
block  at the lower right corner of $h\Phi$ and $Y Y$ 
into blocks of 
size $2n$ and $2(N_c-k-n)$ as follows:
\bea
h\Phi = \left(
\begin{array}{ccc}
0 & 0 & 0  \\
0 & h \Phi_{2n}  & 0 \\
0 & 0 & \frac{\mu^2}{\mu_{\phi}} {\bf 1}_{N_c-k-n}  \otimes  \sigma_3
\end{array}
\right), \qquad
Y  Y = \left(
\begin{array}{ccc}
\mu^2 {\bf 1}_{2k} & 0 & 0  \\
0 &  \varphi  \varphi  &  0 \\
0 & 0 & 0
\end{array}
\right).
\nonu
\eea
Here $\varphi$  is $2n \times 2(\widetilde{N}_c'-k)$
dimensional matrix and corresponds to $2n$ flavors of fundamentals of
the gauge group $SO(2\widetilde{N}_c'-2k)$ which is unbroken by the nonzero
expectation value of $Y$.
In the brane configuration, 
they correspond to 
fundamental strings connecting the $n$ upper flavor D4-branes and
$(\widetilde{N}_c'-k)$
color D4-branes(and their mirrors).
The $\Phi_{2n}$ and $ \varphi \varphi$
are $2n \times 2n$ matrices.
The supersymmetric ground state corresponds to
$h\Phi_{2n}= \frac{\mu^2}{\mu_{\phi}} {\bf 1}_{n}  \otimes  
\sigma_3, 
\varphi =0$. 

The full one loop potential for 
$\Phi_{2n}, \varphi$
takes the similar form in \cite{GK0710-1}
and differentiating this potential with respect to 
$\Phi_{2n}$ and putting $\varphi=0$, one obtains
\bea
h \Phi_{2n} 
\simeq \frac{ \mu_{\phi}^{\ast}}{\widetilde{N}_c'}
{\bf 1}_{n}  \otimes  \sigma_3  \qquad \mbox{or} \qquad
M_{2n} \simeq \frac{\alpha \Lambda^3}{\widetilde{N}_c'} 
{\bf 1}_{n}  \otimes  \sigma_3  
\label{vac2}
\eea
for real $\mu$ and 
we assume here that 
$\mu_{\phi} << \mu << \Lambda_m$. The vacuum energy $V$ is given by
$V \simeq n |h \mu^2|^2$ and
expanding around this solution, one obtains
the eigenvalues for mass matrix for $\varphi$  
and the vacuum (\ref{vac2}) is locally stable.
One can also analyze the correspondence between the eigenvalues of
$h\Phi$ and the $w$ coordinates for the flavor D4-branes.

\section{Meta-stable brane configuration with five NS-branes plus O6-plane}

\subsection{Electric theory}

The type IIA brane configuration \cite{Ahn07-5} corresponding to 
${\cal N}=1$ supersymmetric gauge theory with
gauge group
\bea
SU(N_c) \times SU(N_c')
\label{secondgauge}
\eea
and the symmetric flavor for $SU(N_c)$, the conjugate 
symmetric flavor for $SU(N_c)$,
a bifundamental $X$ in the representation 
$({\bf N_c, \overline{N_c'}})$ and its conjugate field $\widetilde{X}$ 
in the representation $({\bf \overline{N_c}, N_c'})$, 
under the gauge group can be described similarly. 
It consists of 
a middle NS5-brane(012345),
the left $NS5_L$-brane(012345) and the right 
$NS5_R$-brane(012345), the left $NS5_L'$-brane(012389) and the right 
$NS5_R'$-brane(012389), $N_c$-  and $N_c'$-color D4-branes(01236) and an 
$O6^{+}$-plane(0123789).
We take the arbitrary number of color D4-branes with the constraint 
$2N_c' \geq N_c$.
The $O6^{+}$-plane acts as $(x^4,x^5,x^6) \rightarrow
(-x^4,-x^5,-x^6)$ and has RR charge $+ 4$ 
playing the role of $+4$
D6-branes.
The bifundamentals $X$ and $\widetilde{X}$  correspond to 4-4 
strings connecting 
the $N_c$-color D4-branes with $N_c'$-color D4-branes while
the symmetric and conjugate symmetric flavors correspond to 
4-4 strings connecting $N_c$ D4-branes located at negative $x^6$
region
and $N_c$ D4-branes located at positive $x^6$ region \footnote{
See also the relevant works in 
\cite{ILS,BH,BIWW,LLL} for supersymmetric vacua and 
\cite{Ahn07-4,Ahn07-9,Ahn07-6,Ahn07-7,Ahn07-10} for 
nonsupersymmetric vacua in the presence of O6-plane.}.

The middle NS5-brane is located at $x^6=0$ and the $x^6$ 
coordinates for the $NS5_L$-brane, $NS5_L'$-brane, $NS5_R'$-brane and
$NS5_R$-brane are given  by $x^6=-y_2, -y_1, y_1$ and
$x^6=y_2$
respectively. 
The $N_c$ D4-branes 
are suspended between the 
$NS5_L'$-brane(whose $x^6$ coordinate is given by $x^6=-y_1$)  and 
$NS5_R'$-brane(whose $x^6$ coordinate is given by $x^6=y_1$)  while 
the $N_c'$ D4-branes 
are suspended between the $NS5_L$-brane and the $NS5_L'$-brane(and 
moreover they are 
suspended between the $NS5_R'$-brane and the $NS5_R$-brane).
We draw this brane configuration in Figure 5A \cite{Ahn07-5} 
for the vanishing mass
for the bifundamentals \footnote{This is equivalent to the reduced brane
configuration of Figure 1 in 
\cite{Ahn07-4} with particular rotations for the
NS-branes if we ignore all the D6-branes 
completely.}.

\begin{figure}[ht]
   \epsfxsize=5.0in 
\centerline{\epsffile{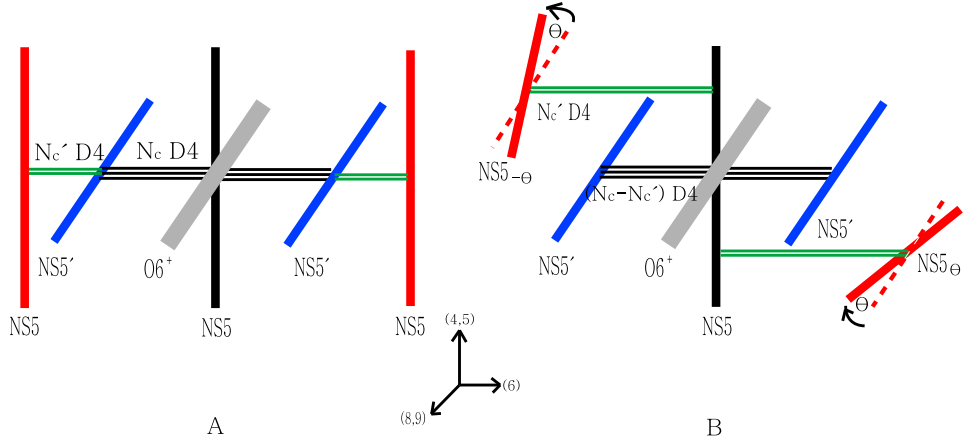}}
   \caption[FIG. \arabic{figure}.]{ 
The  ${\cal N}=1$ supersymmetric 
electric brane configuration for the gauge group $SU(N_c) \times
SU(N_c')$ and the bifundamentals $X$ and $\widetilde{X}$  
as well as symmetric and conjugate symmetric flavors with 
vanishing mass term(5A) and 
nonvanishing mass  and quartic terms(5B) 
for the bifundamentals. 
In Figure 5B, in addition to the mass deformation,
the deformation by the rotation of
$NS5_L$-brane by an angle $-(\frac{\pi}{2}-\theta)$ in $(w,v)$-plane arises.
The superpotential of Figure 5B 
is characterized by (\ref{superelectric2}) and the
mirrors are preserved under the O6-plane
action. }
\end{figure}

The gauge couplings of $SU(N_c)$ and $ SU(N_c')$
are given by
\bea
g_1^2 = \frac{g_s \ell_s}{y_1}, \qquad 
g_2^2 = \frac{g_s \ell_s}{y_2}.
\label{couplings}
\eea
As $y_2$ goes to $\infty$, the $SU(N_c')$ gauge group becomes a
global symmetry and the theory leads to SQCD-like theory with the gauge group
$SU(N_c)$ with symmetric and conjugate symmetric flavors 
and $N_c'$ fundamental flavors.

According to result of \cite{Ahn07-4}, there is no electric
superpotential corresponding to the Figure 5A. Now let us deform 
this theory
\footnote{Another deformation corresponds to a rotation $\theta'$ of the
$NS5_L'$-brane and the $NS5_R'$-brane in $(w,v)$-plane. This
introduces the dynamics of an adjoint of gauge group $SU(N_c)$ whose mass goes
to infinity in the limit where the rotation angle goes to zero  but is
finite for generic nonzero rotation angle $\theta'$. This field couples to the
symmetric tensor $S$, its conjugate field $\widetilde{S}$ 
and bifundamentals $X, \widetilde{X}$. 
Integrating it out leads to 
a further contribution to the quartic supperpotential for
bifundamentals (\ref{superelectric2}). Moreover, there are also 
two terms $(S \widetilde{S})^2$
and $S \widetilde{S} X \widetilde{X}$ with angle-dependent
coefficient in the superpotential. These extra terms appear in the
magnetic superpotential also. This is beyond the scope of the present paper.}. 
Displacing the two NS5-branes relative each other in the $v$ 
direction corresponds to turning on a quadratic
superpotential
for the bifundamentals $X$ and $\widetilde{X}$ and further rotation of
$NS5_L$-brane by an angle $-(\frac{\pi}{2}-\theta)$ 
in $(w,v)$-plane(and $NS5_R$-brane
by an angle $+(\frac{\pi}{2}-\theta)$ in $(w,v)$-plane)
provides the following deformed electric superpotential
\bea
W_{elec} = -\frac{\alpha}{2}   \tr (X \widetilde{X})^2 + m \tr X \widetilde{X},
\qquad
\alpha = \frac{\tan \theta}{\Lambda}, 
\qquad m = \frac{v_{NS5_{-\theta}}}{2\pi \ell_s^2}. 
\label{superelectric2}
\eea
The $NS5_{-\theta}$-brane is moving to the $+v$ direction together
with $N_c'$ D4-branes while the 
$NS5_{\theta}$-brane is moving to $-v$ direction due to the O6-plane for
fixed NS5-brane and $NS5_{L,R}'$-branes.

We draw this brane configuration in Figure 5B for nonvanishing mass
for the bifundamentals
by moving the $NS5_L$-brane with 
$N_c'$ color D4-branes 
to the $+v$ direction(and their mirrors to $-v$
direction) and rotating  it by an angle 
$-(\frac{\pi}{2}-\theta)$ in $(w,v)$-plane. 
Here we decompose $N_c$ D4-branes connecting $NS5_L'$-brane and
$NS5_R'$-brane into $(N_c-N_c')$
D4-branes and $N_c'$ D4-branes. Then 
the latter can move $+v$ direction together with same number of
D4-branes connecting between 
the $NS5_{-\theta}$-brane and the $NS5_L'$-brane(and their mirrors)
while the former are connecting
between the $NS5_L'$-brane and the NS5-brane(and their mirrors). 
Compared with the brane configuration of \cite{Ahn07-5}, the rotation
of $NS5_L$-brane is the only difference and $\theta \rightarrow 
\frac{\pi}{2}$ limit reduces to the one of \cite{Ahn07-5}. 

The solution for the supersymmetric vacua can be obtained by 
$X \widetilde{X} =\frac{m}{\alpha}$ through the F-term conditions.
This breaks 
the gauge group $SU(N_c) \times SU(N_c')$ to $SU(N_c-N_c'-k)$, 
$SU(N_c'-k)$, $SO(2k)$
and $U(k)$.
When the middle NS5-brane moves to $+ w$ direction, then the three
NS-branes intersect in three points in $(v,w)$-plane. 
In other words, the coordinates of $(v,w)$ for those points are
$(v_{NS5_{-\theta}}, v_{NS5_{-\theta}} \cot \theta)$, $(0, 
v_{NS5_{-\theta}} \cot \theta)$ and $(0, 
2v_{NS5_{-\theta}} \cot \theta)$.
It is easy to see that there exists the other intersection point 
given by $(-v_{NS5_{-\theta}}, 
v_{NS5_{-\theta}} \cot \theta)$. Note that the distance along 
$v$ coordinate for 
$NS5_{\theta}$-brane is equal to the one for $NS5_{-\theta}$-brane: 
$v_{NS5_{\theta}} =
v_{NS5_{-\theta}}$.
Then $(N_c-N_c'-k)$
D4-branes are connecting between the $NS5_L'$-brane and the
$NS5_R'$-brane including the middle NS5-brane.
The $(N_c'-k)$
D4-branes are connecting between the $NS5_{-\theta}$-brane and 
the middle NS5-brane(and their mirrors). 
The $k$
D4-branes are connecting between the $NS5_{-\theta}$-brane and the
$NS5_L'$-brane(and their mirrors).
Finally,  
$2k$
D4-branes are connecting between the $NS5_L'$-brane and the
$NS5_R'$-brane directly without touching the middle NS5-brane.
The distance from $2k$ D4-branes to the middle NS5-brane can be read
off from the trigonometric geometry and it is given by
$w=+ v_{NS5_{-\theta}} \cot \theta$. 

\subsection{Magnetic theory}

Starting from the Figure 5B, we apply the Seiberg dual to the 
$SU(N_c)$ factor in (\ref{secondgauge}), the $NS5_L'$-brane and 
the $NS5_R'$-brane are 
interchanged each other. Then the number of color $\widetilde{N}_c$
was given by $\widetilde{N}_c=2N_c'-N_c$ from \cite{Ahn07-4,Ahn07}.
Then the $N_c'$ 
D4-branes are connecting between the $NS5_{-\theta}$-brane and the 
$NS5_R'$-brane(and their mirrors) and 
$\widetilde{N}_c$ D4-branes are connecting between $NS5_R'$-brane and   
the middle NS5-brane. 
By introducing $N_c'$ D4-branes and $N_c'$ 
anti-D4-branes  between the $NS5_R'$-brane and   
the NS5-brane, reconnecting the former with  
the $N_c'$ D4-branes connecting between the $NS5_{-\theta}$-brane 
and the $NS5_R'$-brane and moving those combined D4-branes
to $+v$-direction(and their mirrors to $-v$ direction), 
one gets the final Figure 6A where we are left with 
$(N_c'-\widetilde{N}_c)$ anti-D4-branes between the $NS5_R'$-brane and   
the NS5-brane.
 
\begin{figure}[ht]
   \epsfxsize=5.0in 
\centerline{\epsffile{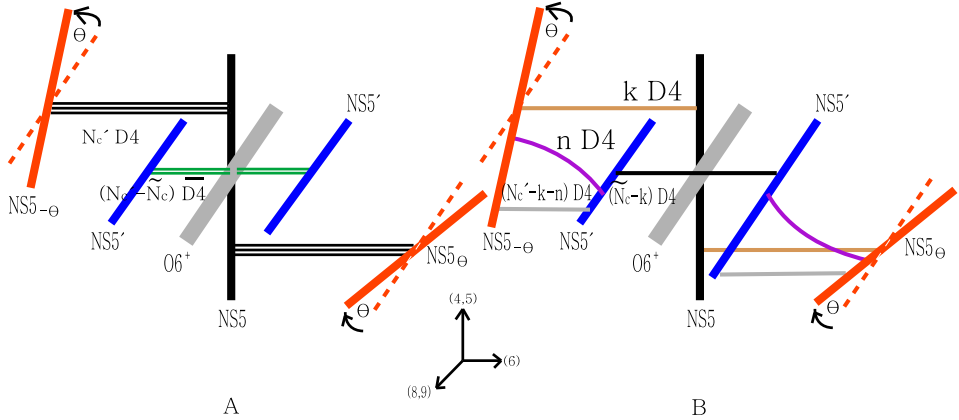}}
   \caption[FIG. \arabic{figure}.]{ 
The  magnetic brane configuration corresponding to Figure 5B with D4-
and $\overline{D4}$-branes(6A) 
when the distance between $NS5_{-\theta}$-brane
and the $NS5_R'$-brane along $v$ direction is large 
and  with 
a misalignment between D4-branes(6B)  when they are close to
each other.
The upper $N_c'$ flavor D4-branes connecting between
the  
$NS5_{-\theta}$-brane and $NS5_R'$-brane are splitting into $(N_c'-k)$ and
$k$ D4-branes. 
The location of intersection between the  $NS5_{-\theta}$-brane
and the upper $(N_c'-k)$
D4-branes 
is given by $(v,w)=(0, +v_{NS5_{-\theta}} 
\cot \theta)$ while the one between  
the $NS5_{-\theta}$-brane and the upper $k$
D4-branes 
is given by $(v,w)=(+v_{NS5_{-\theta}}, 0)$. By moving $n$ flavor D4-branes 
from the upper $(N_c'-k)$ flavor D4-branes,  
the nonzero positive 
$w$ coordinate for $n$  ``curved'' flavor D4-branes can be
determined.
Similarly, the location of intersection between the $NS5_{\theta}$-brane
and the lower $(N_c'-k)$
D4-branes 
is given by $(v,w)=(0, +v_{NS5_{-\theta}} 
\cot \theta)$ while the one between  
the  $NS5_{\theta}$-brane and the lower $k$
D4-branes 
is given by $(v,w)=(-v_{NS5_{-\theta}}, 0)$ due to the O6-plane 
action. }
\end{figure}

The gauge group is given by
\bea
SU(\widetilde{N}_c=2N_c'-N_c) \times SU(N_c')
\label{dual2}
\eea
where the number of dual color can be obtained from the linking number 
counting, as done in \cite{Ahn07-4,Ahn07}.
The matter contents are the field $Y$ in the bifundamental 
representation $({\bf \widetilde{N}_c, \overline{N_c'}})$
and its complex conjugate field $\widetilde{Y}$ in the bifundamental 
representation  $({\bf \overline{\widetilde{N}_c}, N_c'})$,
under the dual gauge
group (\ref{dual2}) 
and  the gauge singlet $M \equiv X \widetilde{X}$ in the representation for 
$({\bf 1,N_c^{'2}-1}) \oplus ({\bf 1, 1})$ under the dual gauge group.
There are also
the symmetric flavor for $SU(\widetilde{N}_c)$ and the conjugate 
symmetric flavor for $SU(\widetilde{N}_c)$.
A cubic superpotential is an interaction between dual ``quarks''
and a meson. 

Then the dual magnetic superpotential, by adding the mass term
like as (\ref{superelectric2}) and the quartic term 
for the bifundamentals 
$X$ and $\widetilde{X}$ to this cubic
superpotential, is given by
\bea
W_{dual} =  \frac{1}{\Lambda} M Y \widetilde{Y}- \frac{\alpha}{2} M^2  + m M. 
\label{superpo1}
\eea
The brane configuration for 
 zero mass for the bifundamentals
can be obtained from Figure 6A by  pushing the  $NS5_{-\theta}$-brane  
together with $N_c'$ D4-branes into the origin $v=0$.
Then the number of dual colors for D4-branes 
becomes $N_c'$ between the $NS5_{-\theta}$-brane and the $NS5_R'$-brane
and $\widetilde{N}_c$ between the $NS5_R'$-brane and the NS5-brane(and
their mirrors).
Note that there is no interaction term between the symmetric or
conjugate symmetric flavors with other matter contents. During the
dual process, the outer $NS5_{\pm \theta}$-branes do not cross the middle 
NS5-brane.

The conditions
  $b_{SU(\widetilde{N}_c)}^{mag} < 0$ and $b_{SU(N_c)} > 0$ imply
that $N_c' < \frac{2}{3} N_c +\frac{2}{3}$. 
Then the range for the $N_c'$ 
can be written as $ \frac{1}{2} N_c < N_c' 
< \frac{2}{3} N_c +\frac{2}{3}$. 
Moreover, $b_{SU(N_c')}= 3N_c'-N_c > 0$ and 
$b_{SU(N_c')}^{mag}=N_c > 0$.
Then 
one can also analyze the
hierarchy of scales as before. 

The brane configuration in Figure 6A is stable as long as the
distance $v_{NS5_{-\theta}}$ between the upper $NS5_{-\theta}$-brane and 
the $NS5_R'$-brane is large, as
in \cite{GK,Ahn07-5}. If they are close to each other, then this brane
configuration is unstable to decay and leads to 
the brane configuration in Figure
6B.
One can regard these brane configurations as particular states in the
magnetic gauge theory with the gauge group (\ref{dual2}) and
superpotential (\ref{superpo1}).
When the $NS5_{-\theta}$-brane 
is replaced by coincident
$N_c'$ D6-branes,
the brane configuration of Figure 6B is the same as the Figure 3 studied in 
\cite{Ahn07-11}. This brane configuration also can be obtained from
the Figure 2B by adding O6-plane with appropriate mirrors.

One can solve the F-term equations (\ref{Fterm})
and one takes
$N_c' \times N_c'$ matrix in (\ref{M0}). 
In the brane configuration of Figure 6B, the $k$ of the
$N_c'$ flavor D4-branes are connecting with $k$ of $\widetilde{N}_c$ color
D4-branes
and the resulting D4-branes stretch from the $NS5_{\theta}$-brane to
the NS5-brane and the coordinate of an intersection point between the 
$k$ D4-branes and the NS5-brane is given by $(v,
w)=(-v_{NS5_{\theta}}, 0)$ where $v_{NS5_{\theta}}$ is a distance
between $v=0$ and the $v$ coordinate of $k$ D4-branes.
This corresponds to the $k$'s eigenvalues $0$ of $M$ (\ref{M0}).
The remaining $(N_c'-k)$ flavor D4-branes between 
the $NS5_{\theta}$-brane and 
the $NS5_L'$-brane are related to the corresponding eigenvalues 
of $M$ (\ref{M0}), i.e.,  $\frac{m}{\alpha} {\bf 1}_{N_c'-k}$.
The coordinate of an intersection point between the 
$(N_c'-k)$ D4-branes and the $NS5_L'$-brane is given 
by $(v, w)=(0, +v_{NS5_{\theta}} \cot \theta)$.
The product $Y \widetilde{Y}$
is given by 
(\ref{solqq})
and 
the supersymmetric vacuum drawn in Figure 6B with $k=0$ has 
the vacuum expectation values $Y  =  \widetilde{Y} =0$ and the gauge group 
$SU(\widetilde{N}_c)$ is unbroken. The expectation value of $M$ 
is given by
$M = \frac{m}{\alpha} {\bf 1}_{N_c'}= m \Lambda \cot 
\theta {\bf 1}_{N_c'}$.


Then the magnetic superpotential can be written in terms of $\Phi$
through (\ref{Dual})
and the classical supersymmetric vacua are given similarly.
Now one splits
the $(N_c'-k) \times (N_c'-k)$
block  at the lower right corner of $h\Phi$ and $Y
\widetilde{Y}$ 
into blocks of 
size $n$ and $(N_c'-k-n)$ as before.
The full one loop potential for 
$\Phi_n, \varphi, \widetilde{\varphi}$
can be described and 
the vacuum expectation value for $h \Phi_n$ 
is given by (\ref{vac}) with an appropriate number of color D4-branes.
One obtains
the eigenvalues for mass matrix for $\varphi$ and 
$\widetilde{\varphi}$ 
and the vacuum (\ref{vac}) is locally stable.
The $n$ flavor D4-branes of 
straight brane configuration
of
Figure 6B can bend due to the fact that there exists an attractive
gravitational interaction
between those flavor D4-branes and the NS5-brane from the DBI action. 

One can move $n$ D4-branes, from $(N_c'-k)$ D4-branes stretched
between the $NS5_L'$-brane and 
the $NS5_{\theta}$-brane at $w=+v_{NS5_{\theta}}
\cot \theta $, to the local minimum of the potential and the end
points of these $n$ D4-branes are at a nonzero $w$ as in Figure 6B.
The remaining $(N_c'-k-n)$ flavor D4-branes between 
the $NS5_{\theta}$-brane and 
the $NS5_L'$-brane are related to the corresponding eigenvalues 
of $h\Phi$ (\ref{Eigen}), i.e.,  
$\frac{\mu^2}{\mu_{\phi}} {\bf 1}_{N_c'-k-n}$.
The coordinate of an intersection point between the 
$(N_c'-k-n)$ D4-branes and the $NS5_L'$-brane is given 
by $(v, w)=(0, +v_{NS5_{\theta}} \cot \theta)$.
The $k$ D4-branes stretching from the $NS5_{\theta}$-brane to
the NS5-brane 
correspond to  exactly the $k$'s eigenvalues 
$0$ of $h\Phi$ (\ref{Eigen}).
Finally, 
the remnant $n$ ``curved'' flavor D4-branes between 
the $NS5_{\theta}$-branes and 
the $NS5_L'$-brane are related to the corresponding eigenvalues 
(\ref{vac}) 
of $h\Phi_n$. It can be computed that
the local minimum occurs at $w \simeq \tan \theta \frac{y^4}{\ell_s^2 
v_{NS5_{\theta}}}$ with $x^6 \equiv y$ \cite{GK0710-1}.

\subsection{Other electric and magnetic theories }


The type IIA brane configuration \cite{Ahn07-5} corresponding to 
${\cal N}=1$ supersymmetric gauge theory with
gauge group (\ref{secondgauge})
and the antisymmetric flavor for $SU(N_c)$, the conjugate 
symmetric flavor for $SU(N_c)$, eight fundamentals for $SU(N_c)$,
a bifundamental $X$ in the representation 
$({\bf N_c, \overline{N_c'}})$ and its conjugate field $\widetilde{X}$ 
in the representation $({\bf \overline{N_c}, N_c'})$, 
under the gauge group can be described similarly:
a middle $NS5_M'$-brane,
the left $NS5_L'$-brane, the right 
$NS5_R'$-brane, the left $NS5_L$-brane, the right 
$NS5_R$-brane, $N_c$-, $N_c'$-color D4-branes,
eight semi-infinite D6-branes, an 
$O6^{+}$-plane and an $O6^{-}$-plane.
We take the arbitrary number of color D4-branes with the constraint 
$2N_c' \geq N_c-4$.
The bifundamentals $X$ and $\widetilde{X}$  correspond to 4-4 
strings connecting 
the $N_c$-color D4-branes with $N_c'$-color D4-branes while
the antisymmetric and conjugate symmetric flavors correspond to 
4-4 strings connecting $N_c$ D4-branes located at negative $x^6$
region
and $N_c$ D4-branes located at positive $x^6$ region.

The middle NS5'-brane is located at $x^6=0$ and the $x^6$ 
coordinates for the $NS5_L'$-brane, $NS5_L$-brane, $NS5_R$-brane and
$NS5_R'$-brane are given  by $x^6=-y_2, -y_1, y_1$ and
$x^6=y_2$
respectively. 
The $N_c$ D4-branes 
are suspended between the 
$NS5_L$-brane(whose $x^6$ coordinate is given by $x^6=-y_1$)  and 
$NS5_R$-brane(whose $x^6$ coordinate is given by $x^6=y_1$)  while 
the $N_c'$ D4-branes 
are suspended between the $NS5_L'$-brane and the 
$NS5_L$-brane(and further they are 
suspended between the $NS5_R$-brane and the $NS5_R'$-brane).
We draw this brane configuration in Figure 7A \cite{Ahn07-5} 
for the vanishing mass
for the bifundamentals \footnote{See also 
the relevant previous work appeared in 
\cite{LLL1,BHKL,EGKT} for supersymmetric vacua and
\cite{Ahn07-4,Ahn07-9,Ahn07-10} for nonsupersymmetric vacua. 
This brane realization is equivalent to the reduced brane
configuration described in section 4 of \cite{Ahn07-4} 
with particular rotations for the
NS-branes if we remove all the D6-branes 
completely.}.
The gauge couplings of $SU(N_c)$ and $ SU(N_c')$
are given by (\ref{couplings}), as before.

\begin{figure}[ht]
   \epsfxsize=5.0in 
\centerline{\epsffile{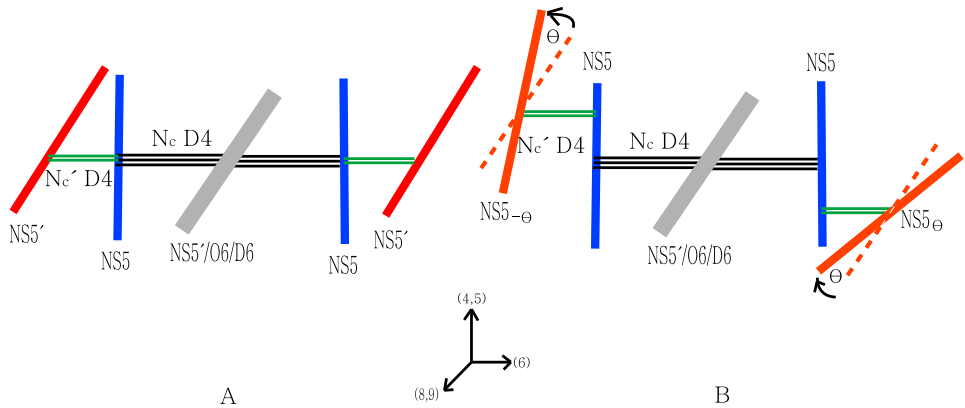}}
   \caption[FIG. \arabic{figure}.]{ 
The  ${\cal N}=1$ supersymmetric 
electric brane configuration for the gauge group $SU(N_c) \times
SU(N_c')$ and the bifundamentals $X$ and $\widetilde{X}$  
as well as antisymmetric, conjugate symmetric flavors and eight D6-branes 
with vanishing mass term(7A) and 
nonvanishing mass and quartic terms(7B) 
for the bifundamentals. In Figure 7B, 
the new deformation by the rotation of
$NS5_L'$-brane by an angle $-\theta$ in $(w,v)$-plane arises(and its
mirror by an angle $\theta$.). }
\end{figure}

There is no electric
superpotential corresponding to the Figure 7A except the interaction
term between the eight fundamental flavors and conjugate symmetric flavor. 
Now let us deform 
this theory. 
Displacing the two NS5'-branes relative each other in the $v$ 
direction corresponds to turning on a quadratic
superpotential
for the bifundamentals $X$ and $\widetilde{X}$.
The $NS5_L'$-brane is moving to the $+v$ direction(and the 
$NS5_R'$-brane is moving to $-v$ direction) due to the O6-plane for
fixed NS5-branes.
Moreover, the rotation of 
the $NS5_{L}'$-brane with an angle $-\theta$ in $(w,v)$-plane
leads to a quartic term for the bifundamentals. 
Then the deformed electric superpotential is given by
(\ref{superelectric2}) as well as an insertion of interaction term 
 between the eight fundamental flavors and conjugate symmetric 
flavor \cite{Ahn08-1}.

We draw this brane configuration in Figure 7B for nonvanishing mass
for the bifundamentals
by moving the $NS5_L'$-brane with 
$N_c'$ color D4-branes 
to the $+v$ direction(and their mirrors to $-v$
direction) and rotating  it by an angle $-\theta$ in $(w,v)$-plane. 
Compared with the brane configuration of \cite{Ahn07-5}, the rotation
of $NS5_L'$-brane is the only difference and $\theta \rightarrow 0$ 
limit reduces to the one of \cite{Ahn07-5}. 

The solution for the supersymmetric vacua can be obtained by 
$X \widetilde{X} =\frac{m}{\alpha}$ through the F-term conditions.
This breaks 
the gauge group $SU(N_c) \times SU(N_c')$ to $SU(N_c-k)$, 
$SU(N_c'-k)$, and $U(k)$.
When the $NS5_L$-brane moves to $+ w$ direction, then the three
NS-branes intersect in three points in $(v,w)$-plane. 
The coordinates of $(v,w)$ for those points are
$(v_{NS5_{-\theta}}, v_{NS5_{-\theta}} \cot \theta)$, $(0, 
v_{NS5_{-\theta}} \cot \theta)$ and $(0, 
2v_{NS5_{-\theta}} \cot \theta)$.
There exists the other intersection point 
given by $(-v_{NS5_{-\theta}}, 
v_{NS5_{-\theta}} \cot \theta)$. 
Then $(N_c-k)$
D4-branes are connecting between the $NS5_L$-brane and the
$NS5_R$-brane.
The $(N_c'-k)$
D4-branes are connecting between the $NS5_{-\theta}$-brane and 
the $NS5_L$-brane(and their mirrors). 
Finally,  
$k$
D4-branes are connecting between the $NS5_{-\theta}$-brane and the
$NS5_{\theta}$-brane directly.
The distance from $k$ D4-branes to the $NS5_L$-brane can be read
off from the trigonometric geometry and it is given by
$w=+ v_{NS5_{-\theta}} \cot \theta$. 


Let us apply the Seiberg dual to the $SU(N_c)$ factor.
Starting from Figure 7B and moving the $NS5_L$-brane to the right all the
way past the $NS5_M'$-brane(and $NS5_R$-brane to the left of $NS5_M'$-brane),
one obtains the Figure 8A.
By introducing $N_c'$ D4-branes and $N_c'$ 
anti-D4-branes  between $NS5_R$-brane and   
$NS5_M'$-brane, 
we are left with 
$(N_c'-\widetilde{N}_c)$ anti-D4-branes between $NS5_R$-brane and   
$NS5_M'$-brane(and its mirrors).
 
\begin{figure}[ht]
   \epsfxsize=5.0in 
\centerline{\epsffile{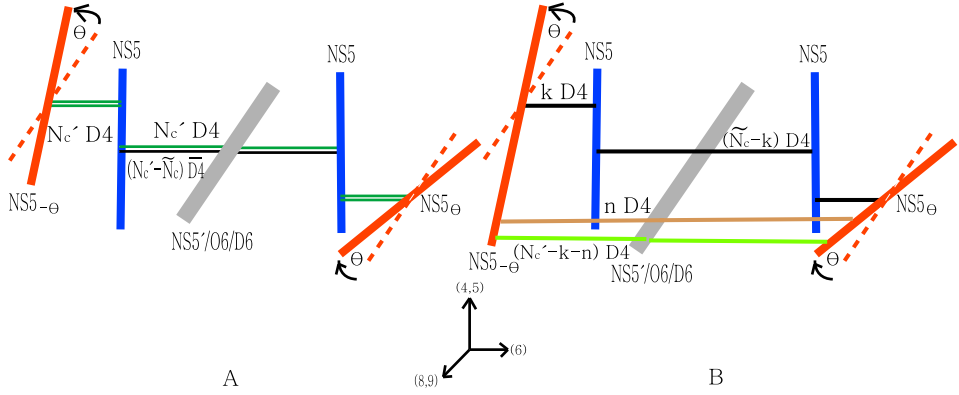}}
   \caption[FIG. \arabic{figure}.]{ 
The  magnetic brane configuration corresponding to Figure 7B with D4-
and $\overline{D4}$-branes(8A) when 
the distance between $NS5_{-\theta}$-brane
and the NS5'-brane along $v$ direction is large 
 and  with 
a misalignment between D4-branes(8B)  when they are close to
each other. 
The upper $N_c'$ flavor D4-branes are splitting into $(N_c'-k)$ and
$k$ D4-branes. 
The location of intersection between the  $NS5_{-\theta}$-brane
and the upper $(N_c'-k)$
D4-branes 
is given by $(v,w)=(0, +v_{NS5_{-\theta}} 
\cot \theta)$ while the one between  
the $NS5_{-\theta}$-brane and the upper $k$
D4-branes 
is given by $(v,w)=(+v_{NS5_{-\theta}}, 0)$. 
By moving $n$ flavor D4-branes 
from the upper $(N_c'-k)$ flavor D4-branes,  
the nonzero positive 
$w$ coordinate for $n$  ``curved'' flavor D4-branes can be
determined.
Similarly, the location of intersection between the $NS5_{\theta}$-brane
and the lower $(N_c'-k)$
D4-branes 
is given by $(v,w)=(0, +v_{NS5_{-\theta}} 
\cot \theta)$ while the one between  
the  $NS5_{\theta}$-brane and the lower $k$
D4-branes 
is given by $(v,w)=(-v_{NS5_{-\theta}}, 0)$ according to 
O6-plane action.}
\end{figure}

The gauge group is given by
\bea
SU(\widetilde{N}_c=2N_c'-N_c+4) \times SU(N_c')
\label{dualdual}
\eea
where the number of dual color can be obtained from the linking number 
counting, as done in \cite{Ahn07-4,Ahn07-1}.
The matter contents are the field $Y$ in the bifundamental 
representation $({\bf \widetilde{N}_c, \overline{N_c'}})$
and its complex conjugate field $\widetilde{Y}$ in the bifundamental 
representation  $({\bf \overline{\widetilde{N}_c}, N_c'})$,
and  the gauge singlet $ X \widetilde{X}$ 
in the representation for 
$({\bf 1, N_c^{'2}-1}) \oplus ({\bf 1, 1})$, under the dual gauge group.
There are also
the antisymmetric flavor $a$, the conjugate 
symmetric flavor $\widetilde{s}$ and eight fundamentals $\hat{q}$ for 
$SU(\widetilde{N}_c)$.

Then the dual magnetic superpotential, by adding the mass term
and quartic term for the bifundamentals $X$ and $\widetilde{X}$, 
is given by
\bea
W_{dual} =  \frac{1}{\Lambda} M Y \widetilde{Y}- \frac{\alpha}{2} M^2
 + m M + \hat{q}
 \widetilde{s} \hat{q}. 
\label{super}
\eea
The brane configuration in Figure 8A is stable as long as the
distance $v_{NS5_{-\theta}}$ between the upper $NS5_{-\theta}$-brane and 
the middle $NS5_M'$-brane is large. If they are close to 
each other then this brane
configuration is unstable to decay and it becomes 
the brane configuration in Figure
8B.
One can regard these brane configurations as particular states in the
magnetic gauge theory with the gauge group (\ref{dualdual}) and
superpotential (\ref{super}).
When the $NS5_{-\theta}$-brane
is replaced by $N_c'$ coincident
D6-branes,
the brane configuration of Figure 8B is the same as the Figure 3 studied in 
\cite{Ahn08-1}. This brane realization can be obtained from the Figure
2B by adding O6-plane with appropriate mirrors.

The conditions
  $b_{SU(\widetilde{N}_c)}^{mag} < 0$ and $b_{SU(N_c)} > 0$ lead to
$N_c' < \frac{2}{3} N_c -\frac{4}{3}$. 
Then the range for the $N_c'$ 
can be written as $ \frac{1}{2} N_c-2 < N_c' 
< \frac{2}{3} N_c -\frac{4}{3}$. 
Moreover, $b_{SU(N_c')}= 3N_c'-N_c > 0$ and 
$b_{SU(N_c')}^{mag}=N_c-4 > 0$.
Then 
one can also analyze the
hierarchy of scales previously.

In the brane configuration of Figure 8B, the $k$ of the
$N_c'$ flavor D4-branes are connecting with $k$ of $\widetilde{N}_c$ color
D4-branes
and the resulting D4-branes stretch from the $NS5_{\theta}$-brane to
the $NS5_L$-brane and the coordinate of an intersection point between the 
$k$ D4-branes and the $NS5_L$-brane is given by $(v, w)=(-v_{NS5_{\theta}}, 0)$.
The coordinate of an intersection point between the 
$(N_c'-k)$ D4-branes and the NS5'-brane is given 
by $(v, w)=(0, +v_{NS5_{\theta}} \cot \theta)$.


Then the magnetic superpotential can be written in terms of $\Phi$
through 
\bea
W_{dual} = 
 h \Phi  Y   \widetilde{Y} 
 +  
\frac{h^2 \mu_{\phi}}{2} \tr \Phi^2- h \mu^2 \tr \Phi 
+ \hat{q}
 \widetilde{s} \hat{q}
\nonu
\eea
and the classical supersymmetric vacua are given similarly.
The full one loop potential for 
$\Phi_n, \varphi, \widetilde{\varphi}$
can be described and 
the vacuum expectation value for $h \Phi_n$ 
is given by (\ref{vac}).
One obtains
the eigenvalues for mass matrix for $\varphi$ and 
$\widetilde{\varphi}$ 
and the vacuum (\ref{vac}) is locally stable.

One can move $n$ D4-branes, from $(N_c'-k)$ D4-branes stretched
between the NS5'-brane and the $NS5_{\theta}$-brane at 
$w=+v_{NS5_{\theta}}
\cot \theta $, to the local minimum of the potential and the end
points of these $n$ D4-branes are at a nonzero $w$ as in Figure 8B.
The coordinate of an intersection point between the 
$(N_c'-k-n)$ D4-branes and the NS5'-brane is given 
by $(v, w)=(0, +v_{NS5_{\theta}} \cot \theta)$.
The $k$ D4-branes stretching from the $NS5_{\theta}$-brane to
the $NS5_L$-brane 
correspond to  exactly the $k$'s eigenvalues 
$0$ of $h\Phi$.
The local minimum occurs at $w \simeq \tan \theta \frac{y^4}{\ell_s^2 
v_{NS5_{\theta}}}$ with $x^6 \equiv y$ \cite{GK0710-1} as before.

\section{Conclusions and outlook}

By following the spirit of \cite{GK0710-1,GK0710}, 
when the quartic term for the bifundamentals 
in the superpotential is present, 
we have constructed the meta-stable brane configurations 
by rotating the NS-brane in $(w,v)$-plane from type IIA string theory. 
They are summarized by the Figures 2B, 4B,
6B and 8B.

As suggested in \cite{GK0710-1}, it would be interesting to study 
what happens 
when we replace the NS5'-brane with other coincident D6-branes further. 
According to the discussion of \cite{GK0710}, there exists a
deformation from higher order terms for the quarks and it would be
interesting to see how these terms can appear in the meta-stable brane
configuration from type IIA string theory. 
It is an open problem to see how to construct a direct gauge mediation for
our meta-stable vacua in the context of \cite{HM,XY}.    

\vspace{.7cm}

\centerline{\bf Acknowledgments}

This work was supported by grant No.
R01-2006-000-10965-0 from the Basic Research Program of the Korea
Science \& Engineering Foundation.  
I would like to thank KIAS(Korea Institute for 
Advanced Study) for hospitality  where
this work was undertaken.

\end{document}